\documentclass[useAMS, usenatbib]{mn2e}
\usepackage{times}
\usepackage{aas_macros}
\usepackage{hyperref}
\usepackage{tablefootnote}
\usepackage{amsmath}
\usepackage{graphicx, subfigure}
\usepackage{color}
\usepackage{float}
\newcommand{\logg}{log \emph{g}}
\newcommand{\teff}{$T_{\rm{eff}}$}
\newcommand{\prot}{$P_{rot}~$}
\newcommand{\w}{\mathbf{w}}
\newcommand{\wh}{$\hat{\mathbf{w}}_n$}

\newcommand{\ph}{$\hat{P}_n$}

\newcommand{\feh}{[Fe/H]}

\newcommand{\nastero}{310}
\newcommand{\nprecise}{14~}

\newcommand{\nHC}{50~}

\newcommand{\ntotal}{365~}
\newcommand{\ngarcia}{310~}
\newcommand{\subcut}{4.2~}

\newcommand{\gyroa}{0.40}
\newcommand{\aerrp}{0.3}
\newcommand{\aerrm}{0.05}
\newcommand{\gyron}{0.55}
\newcommand{\nerrp}{0.02}
\newcommand{\nerrm}{0.09}
\newcommand{\gyrob}{0.31}
\newcommand{\berrp}{0.05}
\newcommand{\berrm}{0.02}

\newcommand{\U}{8}  
\newcommand{\V}{2.1}  
\newcommand{\W}{9.9}  
\newcommand{\X}{14} 
\newcommand{\Y}{5.0}  
\newcommand{\Z}{16.1}  
\newcommand{\Q}{0.14}
\newcommand{\Uerrp}{4}
\newcommand{\Uerrm}{2}
\newcommand{\Verrp}{1}
\newcommand{\Verrm}{0.6}
\newcommand{\Werrp}{0.7}
\newcommand{\Werrm}{0.5}
\newcommand{\Xerr}{4}
\newcommand{\Yerrp}{1}
\newcommand{\Yerrm}{0.8}
\newcommand{\Zerrp}{0.7}
\newcommand{\Zerrm}{0.8}
\newcommand{\Qerrp}{0.06}
\newcommand{\Qerrm}{0.05}

\title{Calibrating Gyrochronology using Kepler Asteroseismic targets}

\author[R.~Angus \emph{et al.}]{%
    Ruth~Angus,$^1$\thanks{ruth.angus@astro.ox.ac.uk}
    Suzanne~Aigrain,$^1$
    Daniel Foreman-Mackey,$^2$ and
    Amy~McQuillan$^3$ \\
    $^1$Department of Physics, University of Oxford, UK \\
    $^2$Centre for Cosmology and Particle Physics, New York University, New York, NY, USA \\
    $^3$School of Physics and Astronomy, Raymond and Beverly Sackler, Faculty of Exact Sciences, Tel Aviv University, 69978, Tel Aviv, Israel}

\begin{document}

\date{Draft version 2014 November 24}
\maketitle

\begin{abstract}

Among the available methods for dating stars, gyrochronology is a powerful one
because it requires knowledge of only the star's mass and rotation period.
Gyrochronology relations have previously been calibrated using young
clusters, with the Sun providing the only age dependence, and are therefore
poorly calibrated at late ages.
We used rotation period measurements of 310 {\it Kepler} stars with
asteroseismic ages, 50 stars from the Hyades and Coma Berenices clusters and
6 field stars (including the Sun) with precise age measurements to calibrate
the gyrochronology relation, whilst fully accounting for measurement
uncertainties in all observable quantities.
We calibrated a relation of the form $P=A^n\times(B-V-c)^b$, where $P$ is
rotation period in days, $A$ is age in Myr, $B$ and $V$ are magnitudes and
$a$, $b$ and $n$ are the free parameters of our model.
We found $a = 0.40^{+0.3}_{-0.05}$, $b = 0.31^{+0.05}_{-0.02}$ and
$n = 0.55^{+0.02}_{-0.09}$.
Markov Chain Monte Carlo methods were used to explore the posterior probability
distribution functions of the gyrochronology parameters and we carefully
checked the effects of leaving out parts of our sample, leading us to find that
no single relation beween rotation period, colour
and age can adequately describe all the subsets of our data.
The {\it Kepler} asteroseismic stars, cluster stars and local field
stars cannot all be described by the same gyrochronology relation.
The {\it Kepler} asteroseismic stars may be subject to observational biases,
however the clusters show unexpected deviations from the predicted behaviour,
providing concerns for the overall reliability of gyrochronology as a dating
method.

\end{abstract}


\section{Introduction}
\label{intro}
\subsection{Dating methods for field stars}

Many fields of astronomy rely on precise age measurements of Main Sequence
(MS) stars.
Unfortunately, age is a notoriously difficult quantity to measure for these
stars, as observable properties evolve slowly on the MS.
Even with high precision spectroscopic measurements, ages often cannot be
determined accurately to within 20\% \citep{Soderblom2010}.
Some of the most precise age measurements currently available are for stars in
clusters where isochrones can be fitted to a coeval population with a range of
masses, resulting in age measurements with uncertainties often as low as 10\%.
Isochronally derived {\it field} star ages, on the other hand, are much less
precise than this, often having uncertainties of around 50\% or more.
Demand for age estimates of planet-hosting stars is high, but faint stars
observed by {\it Kepler} are often expensive or impractical spectroscopic
targets.
Where high resolution spectra are unavailable, gyrochronology can be
extrememly useful.
Gyrochronology is a dating method that utilises the potentially
predictable rotation period evolution of low mass, MS stars.
It requires only knowledge of the current rotation period---which is often
easily extracted from {\it Kepler} light curves---and mass (or appropriate
proxy) of a star.
However the current gyrochronology relations are entirely empirically
calibrated and still need refining at large stellar ages.
{\it Kepler} provides the perfect opportunity to calibrate gyrochronology at
late ages---it provides surface rotation periods of thousands of stars and new
age estimates for hundreds of stars via asteroseismology \footnote{Note that
asteroseismic ages are only available for those {\it Kepler} stars which display
high signal-to-noise Solar-like oscillations in their power
spectra.
The majority of stars that fall into this category are around the same
temperature as, and slightly hotter than, the Sun}.  
This paper aims to use these new asteroseismic age measurements to improve the
gyrochronology relations at late ages.

\subsection{Gyrochronology}

Mass loss via a magnetised stellar wind causes magnetic braking of MS stars
\citep{Weber1967}.
A dynamo-driven magnetic field, generated at the tachocline (the interface
between radiative and convective zones) locks the stellar wind to the surface
of the star.
The stellar wind corotates with the stellar surface out to the Alfv\'{e}n
radius, at which point it decouples and angular momentum is lost from the
star.
The strength of the magnetic field at the surface, and therefore the rate of
angular momentum loss, is related to rotation period \citep{Kawaler1988}.
Due to this dependence, it is postulated that although stellar populations
are born with a range of rotation periods, the rapid rotators rapidly lose
angular momentum and rotation periods converge onto a well-defined sequence.
The timescale for convergence is expected to be around the age of the Hyades
for solar mass stars: 650 Myrs \citep{Radick1987, Irwin2009}.
After this time, rotation periods are thought to be \emph{independent} of their
initial values.  
Gyrochronology postulates that each star falls on a single three-dimensional
plane described by mass, rotation period and age, i.e. given any two of these
three properties, one can determine the third.
The form of angular momentum evolution described above and calibrated in this
article can only be applied to F, G and K MS stars.
Fully convective M dwarfs have a different dynamo-driven magnetic field: their
rotation periods evolve over extremely long timescales and they often do not
converge onto the mass-period-age plane, even after several Gyrs.
Hot stars with effective temperatures greater than $\sim$ 6250 K have shallow
convective zones---they are almost fully radiative---and, again, they have
a different dynamo-driven magnetic field \citep{Kraft1967}.
These massive stars retain their initial rotation period throughout their
brief MS lifetimes and are therefore not suitable gyrochronology targets.

The rate of rotation period decay for intermediate mass MS stars was first
quantified by \citet{Skumanich1972}, who observed that rotation period,
lithium abundance and chromospheric activity decay was proportional to the
square-root of age.
Later, \citet{Noyes1984_2} added a mass dependence to the period-activity-age
relation after more massive stars were observed to spin down more slowly.
The term `gyrochronology' was coined by \citet{Barnes2003} who proposed an
empirically motivated functional form for the relation between period, colour
and age, \begin{equation} \label{eq:Barnes2007_2} P = A^n \times a(B-V-c)^b,
\end{equation} where $P$ is rotation period (in days), $A$ is age (in Myr),
$B$ and $V$ are B and V band magnitudes respectively and $a$, $b$, $c$ and $n$
are dimensionless free parameters.  

This gyrochronology relation was calibrated using open clusters, which are
invaluable calibration tools since their ages are known relatively precisely
and each cluster contains many stars of the same age which enables the
period-mass relation to be tightly constrained.
Unfortunately however, the majority of nearby clusters are young and until
recently it was difficult to measure rotation periods for all but the
youngest, most active stars (using ground--based observations).
There is a significant dearth of precisely measured ages for old stars and it
is for this reason that the current gyrochronology relations are poorly
calibrated at late ages.
\citet{Barnes2007} used 8 young open clusters aged between 30 and 650 Myrs to
calibrate the dependence of rotation period on mass, and the Sun to calibrate
the dependence on age.
Best-fit values of $n$, $a$ and $b$ ($c$ was fixed at 0.4) reported in
\citet{Barnes2007} are presented in table \ref{tab:constants}.
Equation \ref{eq:Barnes2007_2} was further calibrated by \citet{Mamajek2008}
using updated rotation period and age measurements of stars in open clusters
$\alpha$ Per \citep{Prosser1995}, Pleiades \citep{Prosser1995,
Krishnamurthi1998}, M34 \citep{Meibom2011_M34}, and Hyades \citep[Henry,
private comm.,][]{Radick1987, Radick1995, Prosser1995, Paulson2004}.
Once again, the Sun was used as an age anchor---a single data point specifying
the shape of the period-age relation.
Whereas \citet{Barnes2007} fixed the position of $c$, the `colour
discontinuity' in equation \ref{eq:Barnes2007_2}, at 0.4, \citet{Mamajek2008}
allow it to be a free parameter in their model.
The values of $n$, $a$ and $b$, resulting from their fit are shown in table
\ref{tab:constants}.
In both of these studies a maximum likelihood fitting approach was used.
This method relies on the assumption that uncertainties are Gaussian, which
may not always be the case, and only takes observational uncertainties on the
dependent variable into account.
As described in \textsection \ref{sec:gyro_cal}, we adopt a fitting method
that properly accounts for uncertainties on all three observed variables:
colour, period and age.

\begin{table*}
\caption{Values of a, b, c and n in \citet{Barnes2007} and
    \citet{Mamajek2008} and Angus et al. (2014). \label{tab:constants}}
\begin{tabular}{lccc}
\hline\hline
    Parameter & \citet{Barnes2007} & \citet{Mamajek2008} & Angus et al. (2014) \\
    \hline
    a & $0.7725 \pm 0.011$ & $0.407 \pm 0.021$ & $\gyroa^{+\aerrp}_{-\aerrm}$ \\
    b & $0.601 \pm 0.024$ & $0.325 \pm 0.024$ & $\gyrob^{+\aerrp}_{-\berrm}$\\
    c & $0.4$ & $0.495 \pm 0.010$ & $0.45$ \\
    n & $0.5189 \pm 0.0070$ & $0.566 \pm 0.008$ & $\gyron^{+\nerrp}_{-\nerrm}$\\
\hline
\end{tabular}
\end{table*}

The data used in this article are described in \textsection \ref{sec:data},
our calibration and model fitting process is outlined in \textsection
\ref{sec:gyro_cal} and the results are presented and discussed in \textsection
\ref{sec:results}.

\section{Observations}
\label{sec:data}

The ages of 505 {\it Kepler} dwarfs and subgiants were published by
\citet{Chaplin2014}.
They made use of two global asteroseismic parameters---the average large
frequency separation and the frequency of maximum oscillations power---to
estimate stellar properties, including the ages, with a grid-based approach
that utilised several different search codes coupled to more than ten grids of
stellar evolutionary models.

The ages quoted in \citet{Chaplin2014} come from one of the grid-code
combinations, with uncertainties reflecting the scatter between the different
sets of results.
\citet{Chaplin2014} used two different sets of effective temperatures: one was
derived using an Infra-Red Flux Method (IRFM) calibration
\citep{Casagrande2010, SilvaAguirre2012} and the other from a recalibration of
the SDSS griz filter KIC photometry by \citet{Pinsonneault2012} using Yale
Rotating Stellar Evolution Code (YREC) models \citep{Demarque2004}.
We use the IRFM temperatures since they are less dependent on metallicity,
which is not well constrained for the asteroseismic sample, and their
uncertainites are more conservative, however our analysis is relatively
insensitive to this choice.
87 stars in the asteroseismic catalogue have spectroscopic measurements of
\teff$~$, and [Fe/H].
These precisely measured spectroscopic properties allowed more tightly
constrained ages to be calculated for these 87 stars, which were
incorporated where available.
In order to produce a relation that predicts the age of a star using only
observable properties, we chose to convert \teff$~$to $B-V$ for the
asteroseismic sample using the relation of \citet{Sekiguchi2000}.
This conversion added an extra element of systematic uncertainty to our data
since the metallicities provided for most of the asteroseismic stars are
simply an average value for the field: $-0.2\pm0.3$ dex \citep[see e.g.][]{Silva_Aguirre2011}.  
However, since the age uncertainties dominate this analysis, we do not expect
this to have a significant impact on our results.

The ages quoted in Chaplin et al. (2014) have typical uncertainties of 35\%.
These large uncertainties are the result of the fact that only approximate
inferences can be made on the ages using the global asteroseismic parameters,
neither of which has an explicit dependence on age.
It will be possible, however, to derive more precise ages for a subset of
these stars.
By measuring the frequency of each oscillation mode individually,
not just the global asteroseismic parameters, one can provide much tighter
constraints on ages.
Ages derived from individual oscillation mode measurements can have
uncertainties as small as 10\% \citep{Brown1994, SilvaAguirre2013}.
However, measuring frequencies for individual oscillation modes is a manual
process and can only be applied in the highest signal-to-noise cases.
\citet{Chaplin2014} predict that around 150 of the 505 stars will be suitable
for this individual oscillation mode treatment.
We obtained precise ages for 42 stars from \citet{Metcalfe2014}, modelled with
the Asteroseismic Modeling Portal (AMP), with effective temperatures and
metallicities from \citet{Bruntt2012}.
Of the 42 stars in \citet{Metcalfe2014}, we only incorporate the `simple
stars' (cool dwarfs) into our sample, ignoring the hotter F stars and more
evolved subgiant stars as these are not expected to follow the simple
gyrochronology relation.

The {\it Kepler} light curves of the 505 asteroseismic targets display
quasi-periodic variations on timescales corresponding to the rotational
periods of the stars---flux variations are produced by active regions on the
stellar surface that rotate in and out of view.
Rotation periods for \ngarcia of these stars are published in
\citet{Garcia2014} who used a combination of an autocorrelation function and
wavelet transform method to measure surface rotation.
From the 505 targets in the original sample, \nastero$~$rotation periods were
reliably measured, \nprecise of which have precise asteroseismic ages from AMP
modelling.
All stars in the asteroseismic sample with rotation periods published by
\citet{McQuillan_2014}, also appear in the rotation period catalogue of
\citet{Garcia2014}.
There is excellent agreement between rotation period
measurements where the two catalogues overlap.
Of the 114 stars which appear in both catalogues, only the rotation periods of
4 were not consistent at the 1$\sigma$ level and of these only 1, KIC 4931390,
was inconsistent at greater than 2$\sigma$.
The similarities between the two catalogues is further described in
\citet{Garcia2014}.  

\begin{figure*}
\begin{center}
\includegraphics[width=6in, clip=true]{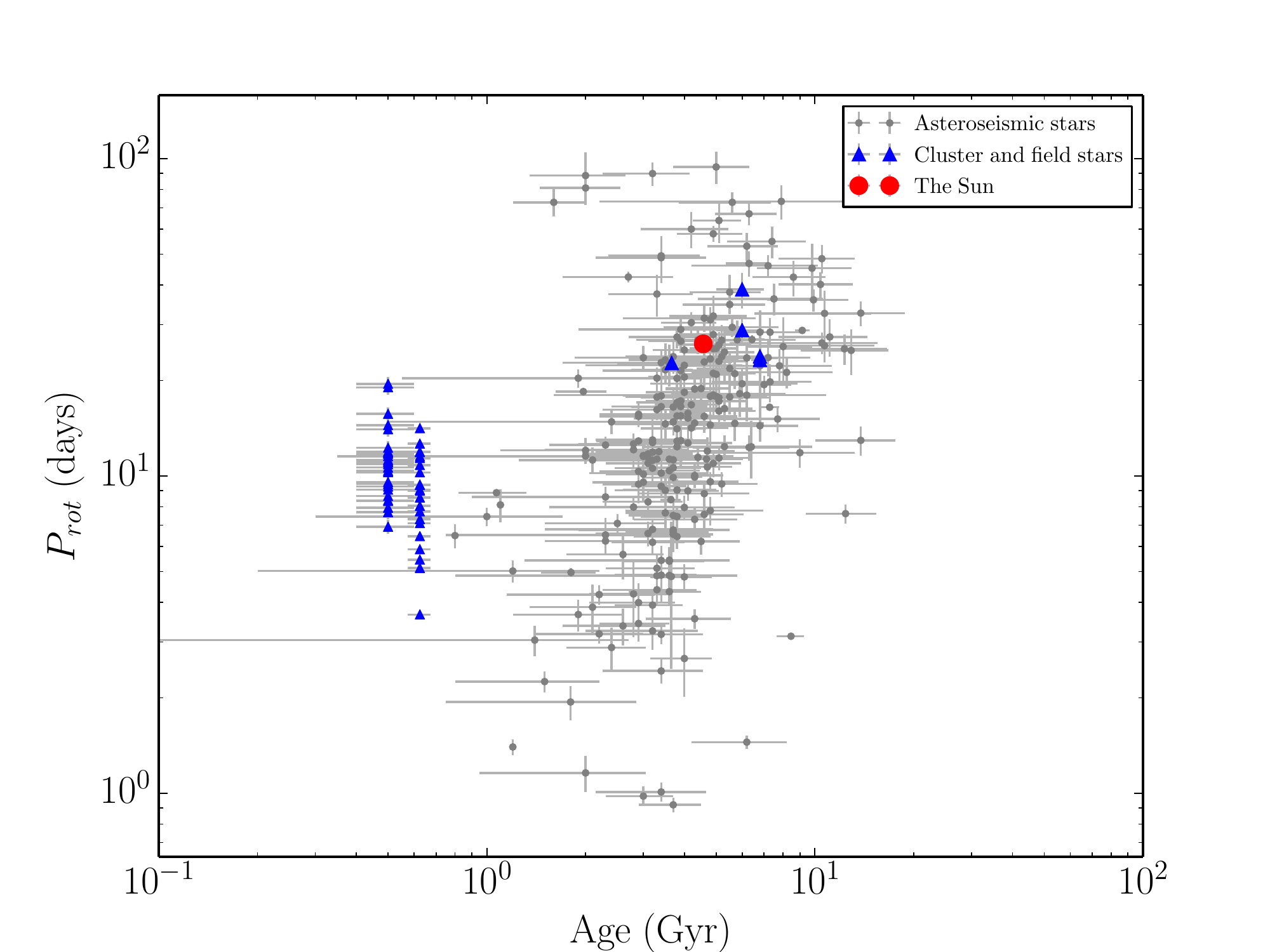}
\caption{Photometric rotation period vs age for \nastero$~$ {\it Kepler}
	targets (grey circles) plus cluster and field stars (blue
	triangles). The Sun is shown as a red circle.
\label{fig:p_vs_a}}
\end{center}
\end{figure*}

\begin{figure*}
\begin{center}
\includegraphics[width=6in, clip=true, trim=0 0 0.5in 0]{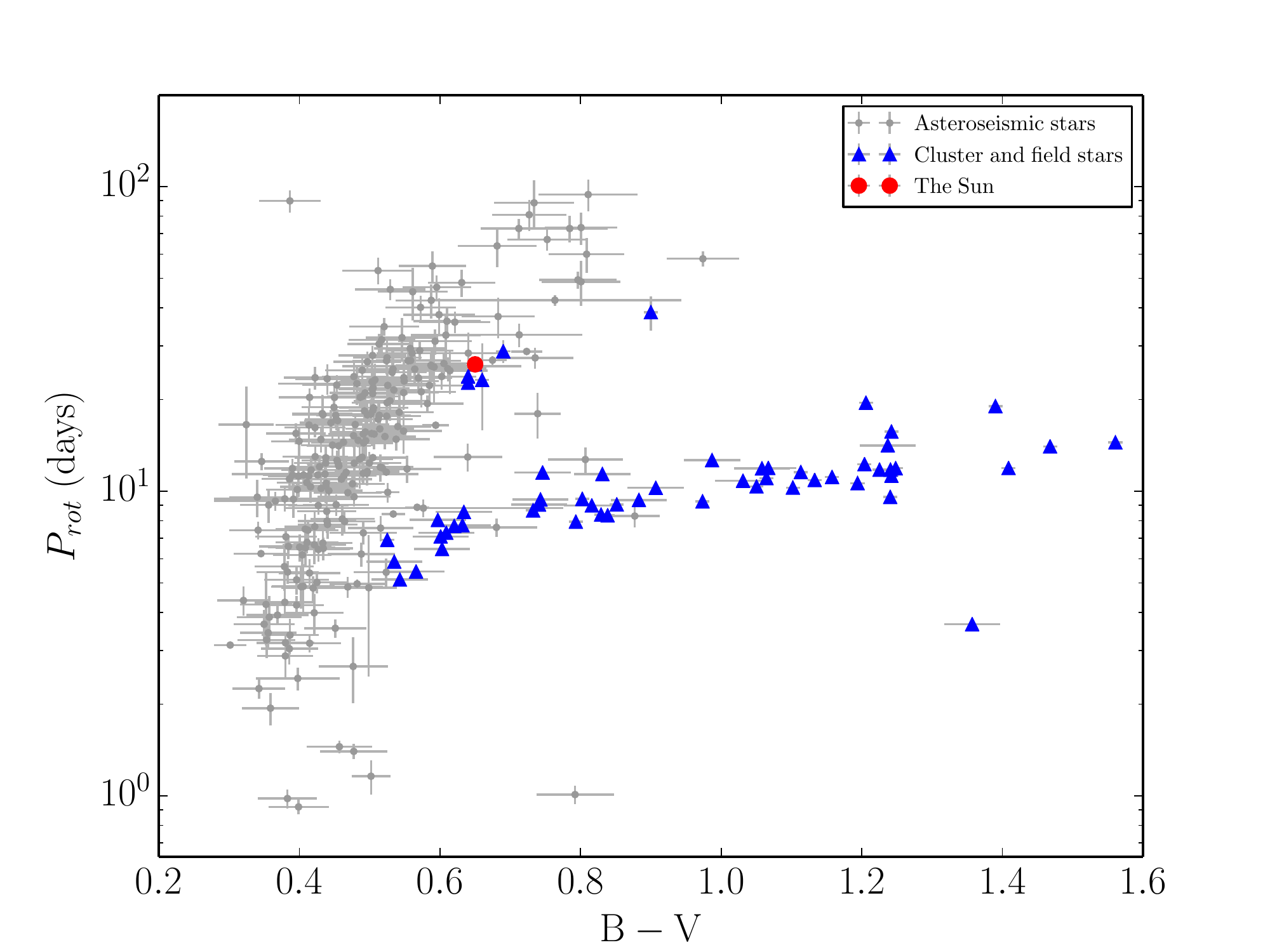}
\caption{Photometric rotation period vs $B-V$ colour for the data described in
	figure \ref{fig:p_vs_a}.
\label{fig:3d}}
\end{center}
\end{figure*}

The asteroseismic sample covers a large range of ages, however it does not
provide good mass coverage across the entire range (see figures
\ref{fig:p_vs_a} and \ref{fig:3d}).
Few stars have temperatures below 6000 K ($B-V$ $\sim$ 0.55) and of the low
mass stars, most of them are old (note that massive stars evolve rapidly and
so we do not expect many in the sample).
The exclusion of young, low-mass stars from the asteroseismic sample is due to
the fact that these stars are more active and their power-spectra do not
display high signal-to-noise acoustic oscillations.
The omission of these as well as other types of stars that are not ideal
asteroseismic targets from our sample should not bias our results.
The mere lack of data in some regions of parameter space will not skew the
best fitting model, however it is important to note that the resulting
gyrochronology relation will not necessarily be descriptive of those stars not represented in this sample.  
We filled in some of the missing parameter space by adding \nHC stars to our
sample from young clusters Coma Berenices (0.5 Gyr), and the Hyades
(0.625 Gyr) (see table \ref{tab:clust}).
Clusters younger than 0.5 Gyr often have large populations of rapid rotators
that have not yet converged onto the gyrochronology plane, so no clusters
younger than Coma Ber were included.
Uncertainties on $B-V$ colours associated with each cluster star were not
provided in the catalogues from which rotation periods and ages were obtained.
Since the uncertainty associated with each measurement plays such a key role
in our analysis (see \textsection \ref{sec:gyro_cal}), we assigned an
uncertainty of
$\pm 0.01$ 
mag to each colour measurement, based on a realistic estimate of the
typical uncertainties expected.
The 1.1 and 0.588 Gyr open clusters, NGC 6811 and Praesepe, were originally
included in our analysis.
However we discovered that their period-colour relations were different to
those of the Hyades and Coma Ber, as well as to each other's, and we therefore
did not include them in our final analysis.
A further 6 field stars with precise age measurements were added to the
sample: 16 Cyg A and B, Alpha Cen A and B, 18 Sco and, of course, the Sun
(see table \ref{tab:field}).
The entire set of \ntotal stars is shown in figures \ref{fig:p_vs_a} and
\ref{fig:3d}.
Asteroseismic targets are shown in grey, with cluster and field stars in blue
and the Sun in red.

\begin{table*}
\caption{Clusters and References: (1) \citet{Dobbie2009},
	(2) \citet{CollierCameron2009}, (3) \citet{Perryman1998},
	(4) \citet{Radick1987}. \label{tab:clust}}
\begin{tabular}{lcccc}
\hline\hline
Cluster & Age (Gyr) & Number of stars & Age ref & Rotation period ref \\
\hline
Coma Ber & 0.5 $\pm$ 0.1 & 28 & 1 & 2 \\
Hyades & 0.625 $\pm$ 0.05 & 22 & 3 & 4 \\
\hline
\end{tabular}
\end{table*}

\begin{table*}
\caption{Rotation periods and $B-V$ colours for field stars with precise
	ages.
Notes: \citet{Davies2014} measured internal rotation periods for 16 Cyg A and B
using asteroseismology.
However, this is likely to be close to the surface rotation value.
Rotation periods for and $\alpha$ Cen A and B were measured
from variation in chromospheric emission lines.
High-resolution spectropolarimetric observations were used to measure
a rotation period for 18 Sco.
The age of 16 Cyg AB was measured with asteroseismology.
18 Sco's age measurement was based chiefly on an asteroseismic analysis,
however its rotation period was used as an additional constraint, so the age
estimate is not entirely independent of rotation period for this star.
An age for the $\alpha$ Cen system was estimated from spectroscopic
observations with additional seismic constraints.\label{tab:field}}

\begin{tabular}{lcccc}
\hline\hline
{ID} & {age} & {\prot} & {$B-V$} & References\\
\hline
16 Cyg A & 6.8 $\pm$ 0.4 & 23.8 $^{+1.5}_{-1.8}$ & 0.66 $\pm$ 0.01 &
\citet{Metcalfe2012}, \citet{Davies2014}, \citet{Moffett1979} \\
16 Cyg B & 6.8 $\pm$ 0.4 & 23.2 $^{+11.5}_{-3.2}$ & 0.66 $\pm$ 0.01 &
\citet{Metcalfe2012}, \citet{Davies2014}, \citet{Moffett1979} \\
18 Sco & 3.7 $\pm$ 0.2 & 22.7 $\pm$ 0.5 & 0.64 $\pm$ 0.01 &
\citet{Li2012}, \citet{Petit2008}, \citet{Mermilliod1986} \\
The Sun & 4.568 $\pm$ 0.001 & 26.09 $\pm$ 0.1 & 0.65 $\pm$ 0.001 &
\citet{Bouvier2010}, \citet{Donahue1996}, \citet{Cox2000} \\
$\alpha$ Cen A & 6 $\pm$ 1 & 28.8 $\pm$ 2.5 &
0.69 $\pm$ 0.01 &
\citet{Bazot2012}, \citet{Yildiz2007}, \citet{Hallam1991}, \citet{Mermilliod1986}\\
$\alpha$ Cen B & 6 $\pm$ 1 & 38.7 $\pm$ 5.0 &
0.90 $\pm$ 0.01 &
\citet{Bazot2012}, \citet{Yildiz2007}, \citet{Dumusque2012}, \citet{Mermilliod1986} \\
\hline
\end{tabular}
\end{table*}

\section{Calibrating the Gyrochronology relation}
\label{sec:gyro_cal}

\subsection{The model}

The \nastero$~$asteroseismic stars in our sample have $B-V$ colours converted
from effective temperatures, photometric rotation periods,
asteroseismic ages, and asteroseismic surface gravities.
Each measurement of these properties is assumed to be independent with an
associated Gaussian uncertainty.
Not all of the cluster stars added to our sample have \logg$~$values; however,
since we only use \logg$~$to separate the populations of subgiants and dwarfs
(and we assume that the cluster stars are dwarfs) this should not affect our
results.
Following the treatment of \citet{Barnes2007} and \citet{Mamajek2008},
rotation period was treated as the dependent variable throughout the modelling
process.

Hot stars and subgiants follow a different gyrochronology relation to MS
dwarfs.
Stars with effective temperatures above the Kraft-break, $T_{\rm{eff}}
\sim$ 6250 K, \citep{Kraft1967} do not have a thick convective envelope and
cannot support a strong magnetic dynamo, so do not spin down appreciably
during their MS lifetimes.
Subgiants spin down rapidly as they expand due to angular momentum
conservation and thus diverge from the gyrochronological mass-period-age
plane.
The point in their evolution at which they depart, the `gyrochronological MS
turn off', is difficult to define.
Classically, MS turnoff is defined as the hottest point on a star's path on
the HR diagram (before it ascends the giant branch) but theory predicts that
evolving stars begin the process of spinning down relatively slowly after
leaving the classically defined MS \citep{vanSaders2013}.
For this reason we chose a very simple definition of MS turnoff: we defined
a \logg$~$boundary of \subcut to mark the transition between dwarfs and
giants.
We tried a range of boundary values and found that \subcut minimised subgiant
contamination whilst maximising the cool dwarf sample.  
It was also necessary to use a mixture model to account for misclassified
subgiants---without it, subgiant contamination significantly biased the
resulting fit.
We did not exclude hot stars and subgiants from our sample during the
modelling process, we modelled all three populations simultaneously.
This allowed for the fact that stars have some probability mass lying in all
three regimes due to their large observational uncertainties.

\begin{figure*}
\begin{center}
\includegraphics[width=6in, clip=true, trim=0 0 0.5in 0]{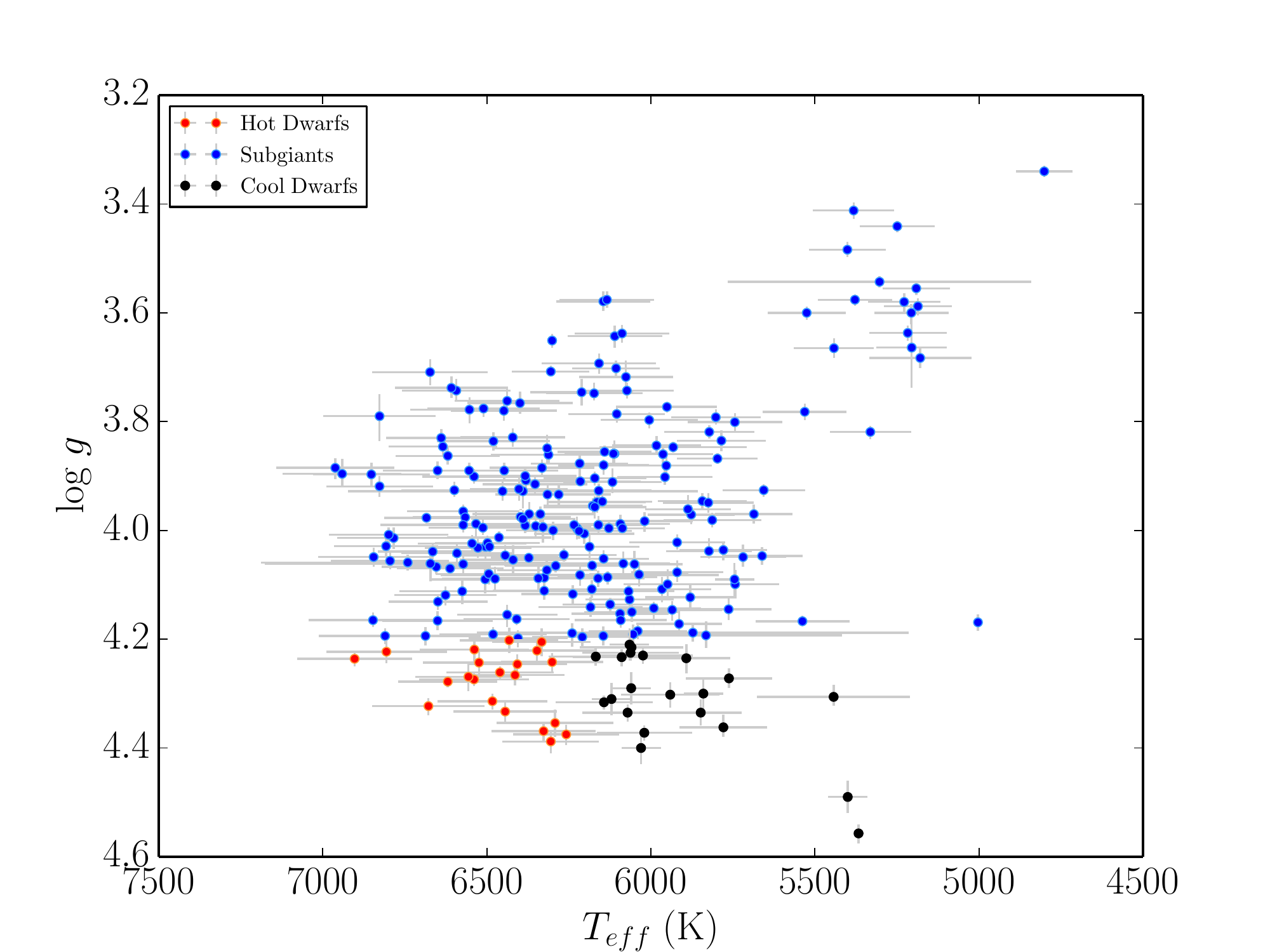}
\caption{\logg$~$vs \teff$~$for the \nastero$~$asteroseismic stars. Hot dwarfs
with \teff$~>$ 6250 K and \logg$~>$ \subcut are red, subgiants with \logg$~<$
\subcut are blue. Only the black cool dwarfs with \teff$~<$ 6250 K and
\logg$~<$ \subcut are expected to follow the gyrochronology relation in
equation \ref{eq:Barnes2007_2}.
\label{fig:logg_vs_t}}
\end{center}
\end{figure*}

Hot MS stars were defined as those with $B-V$ $<$ 0.45, corresponding to
\teff$~\approx$ 6250 K for solar metallicity and \logg.
Since there is no dependence of rotation period on age for massive MS stars,
their rotation periods were modelled as a normal distribution with mean and
standard deviation, $Y$ and $V$, as free parameters.
Subgiant rotation periods \emph{do} depend on age and $T_{\rm{eff}}$.
However, since the rotational properties of these stars are not interesting for
the purposes of gyrochronology calibration, we also modelled them with
a normal distribution with mean and standard deviation, $Z$ and $U$, as free
parameters.
We used a mixture model for the remaining population of stars, consisting of
cool dwarfs and misclassified, contaminating subgiants.
The subgiants were treated as if their rotation periods had been drawn from
a background normal distribution with mean and standard deviation, $X$ and
$U$, again inferred from the data and another parameter, $Q$, the
probability of each star belonging to that background population.
The results of this analysis were not particularly sensitive to the choice of
distribution for the hot stars and subgiants.
These models were put in place so that we did not have to throw any data
away and we could model everything at once.
This was important because stars were classified according to their observed
temperatures and \logg s, which are noisy.
Due to the large uncertainties on \teff$~$and \logg, each star therefore has some
probability of being a subgiant, some of being a cool dwarf and some of
being a hot star.
By throwing away data, one could accidentally throw away a misclassified star.
We avoid this problem by modelling all stars simultaneously and taking a
probabilistic approach to classification.
Inferences made about the parameters of the normal distributions used to model
subgiants and hot stars were not of particular interest for the purposes of
gyrochronology calibration.
We decided to use simple normal distributions rather than more physically
motivated models in order to remain as model--independent as
possible.
Figure \ref{fig:logg_vs_t} shows \logg$~$vs \teff$~$for the asteroseismic stars.
Stars classified as cool dwarfs are shown in black, hot dwarfs in red and
subgiants in blue.

Ideally both the hot star ($B-V$ $<$ 0.45) and subgiant (\logg $<$ \subcut)
boundaries would be free parameters in our model.
However, since these two populations were modelled with a relatively
unconstraining normal distributions, these boundary parameters would not be
well behaved.
Both would be pushed to higher and higher values until all stars were modelled
with a normal distribution.
In order to avoid this problem, we fixed these two boundaries.
A future analysis could avoid the assumption that the gyrochronology relation
is infinitely narrow and assign it some intrinsic width, which would also be
a free parameter.

We postulate that there is a deterministic relationship between the `true'
rotation period of a cool MS star and its `true' age and colour, described by
equation \ref{eq:Barnes2007_2} (by `true' we mean the value an observable
property would take, given extremely high signal-to-noise measurements).
Rotation period also depends on \logg$~$since this property determines whether
a star falls in the dwarf or subgiant regime.

In what follows we use $P$ to denote rotation period and define
$\mathbf{w} = (age, B-V,~\log~g)$ as the vector of additional observational
properties.
Observations are denoted as \ph$~$and $\hat{\mathbf{w}}_n$ and the unobserved
(latent), `true' parameters as $P_n$ and $\mathbf{w}_n$ for stars $1,...,N$.

In order to explore the posterior Probability Distribution Functions (PDFs) of
the model parameters, $\theta$, conditioned on a set of noisy observations,
$\{\hat{P}_n, \hat{\mathbf{w}}_n\}$, it was necessary to marginalise over the
latent parameters, $\{P_n, \mathbf{w}_n\}$.
This is because the model parameters, $\theta$ are conditionally dependent on
the `true' values of rotation period and colour, \emph{not} the observed
values themselves.
The observed values are only conditionally dependent on the `true' values.
In order therefore to infer the values of the model parameters using the
observed values of rotation period and colour, it was necessary to marginalise
over the latent parameters.  
Assuming all measurements are independent, the marginalised likelihood can be
written
\begin{equation}
	p(\{\hat{P}_n,\hat{\w}_n\}|\theta) =
	\prod_{n=1}^{N} \int p(\hat{P}_n,\hat{\w}_n,P_n,\w_n|\theta)
	{\rm d}P_n {\rm d}\w_n.
\label{eq:fulll}
\end{equation}
The joint probability, on the right hand side of this equation, can be
factorised as
\begin{align}
	p(\hat{P}_n,\hat{\w}_n,P_n,\w_n|\theta) = & \\
	p(P_n\,| & \,\w_n,\theta)
	p(\hat{P}_n\,|\,P_n)\,p(\hat{\w}_n\,|\,\w_n)p(\w_n),
\nonumber
\end{align}
where we have utilised the fact that the observations, \ph$~$and \wh$~$are
{\it conditionally independent} of the model parameters, $\theta$: they only
depend on $\theta$ through the latent parameters, $P_n$ and $\w_n$.
The above integral can  be written
\begin{eqnarray}
	p(\{\hat{P}_n,\hat{\w}_n\}|\theta) \propto &
	\prod_{n=1}^{N} \int p(\w_n|\hat{\w}_n) {\rm d}\w_n \\ \nonumber
	& \int p(P_n|\w_n,\theta) p(\hat{P}_n\,|\,P_n) {\rm d}P_n,
\label{eq:fac}
\end{eqnarray}
where we have used Bayes' theorem:
$p(\w_n|\hat{\w}_n) \propto p(\hat{\w}_n|\w_n)p(\w_n)$.
The outer integral is the same for hot dwarfs, cool dwarfs and subgiants
alike.
In our model, the probability of the `true' rotation period given the `true'
observed parameters and the model parameters, $p(P_n|\w_n, \theta)$, is
different in each regime because a different generative process is responsible
for producing rotation periods.
For cool dwarfs ($B-V$ $<$ 0.45 and \logg$~>$ \subcut):
\begin{eqnarray} \label{eq:codw}
p(P_n|\w_n,\theta) &=&
    (1-Q)~\delta \left (P_n - f_\theta(\w_n)\right) \\ \nonumber
    && +\quad Q\,\left(\sqrt{2\pi U^2}\right)^{-1/2} \\ \nonumber
    &&	~\exp\left({-\frac{(P_n-X)^2}{2U^2}}\right),
\end{eqnarray}
where $Q$ is the probability that a star is drawn from the population of
misclassified subgiants and
\begin{eqnarray}
f_\theta(\w_n) = A^n \times a(B-V - c)^b
\end{eqnarray}
is the gyrochronology relation.
For hot dwarfs ($B-V$ $<$ 0.45 and \logg$~>$ \subcut) the generative process
is:
\begin{eqnarray}
p(P_n\,|\,\w_n,\theta) = \left(\sqrt{2\pi V^2}\right)^{-1/2}~
\exp\left({-\frac{(P_n-Y)^2}{2V^2}}\right),
\end{eqnarray}
and for subgiants,
\begin{eqnarray}
p(P_n\,|\,\w_n,\theta) = \left(\sqrt{2\pi W^2}\right)^{-1/2}~
\exp\left({-\frac{(P_n-Z)^2}{2W^2}}\right).
\end{eqnarray}

We used hierarchical inference to account for observational uncertainties,
following the method of \citet{Hogg2010}, also used by
\citet{Foreman-Mackey2014}, \citet{Rogers2014}, \citet{Morton2014} and
\citet{Demory2014}.
We computed equation \ref{eq:fac} up to an unimportant constant
using a sampling approximation.
The values of \ph$~$and \wh$~$with uncertainties, $\sigma_P$ and
$\sigma_{\mathbf{w}}$, reported in catalogues provide constraints on the
posterior probability of those variables, under a choice of prior PDF,
$p_0(\hat{\mathbf{w}}_n)$.
Ideally, these catalogues would provide posterior PDF samples, not just point
estimates, which we could use directly.
i.e. samples from
\begin{equation}
	p(\mathbf{w}_n|\hat{\mathbf{D}}_n) =
	\frac{p(\hat{\mathbf{D}}_n|\mathbf{w}_n)p_0(\mathbf{w}_n)}
	{p_0(\hat{\mathbf{D}}_n)},
\end{equation}
where $p(\hat{\mathbf{D}}_n|\w_n)$ is the likelihood of the data,
$\hat{\mathbf{D}}_n$ (in this case,
the set of {\it Kepler} lightcurves plus spectroscopic \teff$~$and
\feh$~$measurements), given the model parameters, $\mathbf{w}_n$.
$p_0(\hat{\w}_n)$ is an uninformative prior probability distribution
function $p_0(\mathbf{w}_n)$ is an uninformative prior PDF, chosen by the
fitter \citep[][used a flat prior PDF in age and \logg]{Chaplin2014}.
In the absence of posterior PDF samples\footnote{Posterior PDF samples
for asteroseismic parameters are now beginning to be published and will be made
available in future publications.} we generated our own from Gaussian
distributions with means, \wh$~$and standard deviations, $\sigma_{\mathbf{w}}$.
$J$ posterior samples were generated for each star (we used $J$ = 500):
\begin{equation}
\w_n^{(j)} \sim p(\w_n\,|\,\hat{\w}_n),
\end{equation}
and were used to evaluate $p(\mathbf{w}_n|\hat{\mathbf{w}}_n)$ up to a
normalisation constant.
Using these samples we computed the marginalised likelihood for a single
star as follows,
\begin{align}
	p(\hat{P}_n,\hat{\w_n}\,|\,\theta) \approx \frac{1}{J_n}
	\sum_{j=1}^{J_n}p(\hat{P}_n\,|\,\mathbf{w}_n^{(j)},\theta) \quad.
\end{align}
The argument inside this sum is given by the integral
\begin{equation}
p(\hat{P}_n\,|\,\mathbf{w}_n^{(j)},\theta) =
    \int p(P_n|\w_n,\theta) p(\hat{P}_n\,|\,P_n) {\rm d}P_n \quad.
\end{equation}
Assuming that the period uncertainties are Gaussian with mean $\hat{P}_n$
and variance ${\sigma_n}^2$, this integral can be evaluated analytically for
each population.
For example, starting from equation~\ref{eq:codw} the result for the cool
dwarfs is
\begin{eqnarray}
p(\hat{P}_n\,|\,\mathbf{w}_n^{(j)},\theta) &=&
    \frac{1-Q}{\sqrt{2\,\pi\,{\sigma_n}^2}} \exp\left( -
        \frac{\left[\hat{P_n} - f_\theta (\w_n^{(j)}) \right] ^2}
             {2\,{\sigma_n}^2}\right) \nonumber\\
    && +
    \frac{Q}{\sqrt{2\,\pi\,(U^2 + {\sigma_n}^2)}} \exp\left( -
        \frac{[\hat{P_n} - X] ^2}{2\,[U^2 + {\sigma_n}^2]}\right) \quad.
\end{eqnarray}
A similar result can be derived for the other populations.

Finally, using these analytic results for the inner integral, the
marginalised log-likelihood from equation~\ref{eq:fac} becomes
\begin{eqnarray}
	\log p(\{\hat{P}_n,\hat{\w}_n\}\,|\,\theta) \approx & \\ \nonumber
    & \log \mathcal{Z} + \sum_{n=1}^N
	\log \left[ \sum_{j=1}^{J_n}p(\hat{P}_n\,|\,\mathbf{w}_n^{(j)},
\theta) \right ]
\end{eqnarray}
where $\mathcal{Z}$ is an irrelevant normalisation constant.
We used {\tt emcee} \citep{Foreman-Mackey2013}, an affine invariant, ensemble
sampler Markov Chain Monte Carlo (MCMC) algorithm, to explore the posterior
PDFs of the model parameters, $\theta$.
Flat prior PDFs were used for each parameter.
Following the above method, a likelihood was computed as follows:
\begin{itemize}
	\item For each star, $J$ samples were drawn from three normal
		distributions: one in colour, one in age and one in \logg,
		where the means and standard deviations of those distributions
		were the observed values and uncertainties.
		This step was performed just once and the
		following steps were performed for each likelihood evaluation.
	\item For those samples that fell in the cool dwarf regime
		($B-V$ $>$ 0.45 and \logg$~>$ \subcut), model rotation
		periods were both calculated using equation
		\ref{eq:Barnes2007_2} and assigned the value of parameter $X$.
		A Likelihood for the two model rotation periods were then
		evaluated using a Gaussian mixture model
		likelihood function.
	\item For the samples that fell in the hot dwarf ($B-V$ $<$ 0.45 and
		\logg$~>$ \subcut) and subgiant (\logg$~<$ \subcut) regimes,
		likelihoods were calculated by comparing observed rotation
		periods with the model rotation periods for the two
		populations: $Y$ and $Z$.
	\item The total log-likelihood for each star was calculated as the
		sum of the log-likelihoods of each of the $J$ samples.
	\item Finally, the sum of individual star log-likelihoods
		provided the total log-likelihood.
\end{itemize}

\section{Results and Discussion}
\label{sec:results}

\begin{table}
	\caption{Median values of $a$, $b$, $c$ and $n$ for individual clusters
		(see equation \ref{eq:Barnes2007_2}).  
\label{tab:cluster_results}}

\begin{center}
\begin{tabular}{lccc}
\hline\hline
{Parameter} & {Coma Berenices} & {Hyades} \\
\hline
a & $0.417^{+0.08}_{-0.07}$ & $0.312^{+0.04}_{-0.06}$ \\
b & $0.271^{+0.05}_{-0.06}$ & $0.410^{+0.05}_{-0.04}$ \\
n & $0.542 \pm 0.03$ & $0.599^{+0.03}_{-0.02}$ \\
\hline
\end{tabular}
\end{center}
\end{table}

\begin{figure}
\begin{center}
	\subfigure[Rotation period vs $B-V-c$ for the Hyades (top) and Coma
	Berenices (bottom).]{
            \label{fig:CF45}
	    \includegraphics[width=\columnwidth]{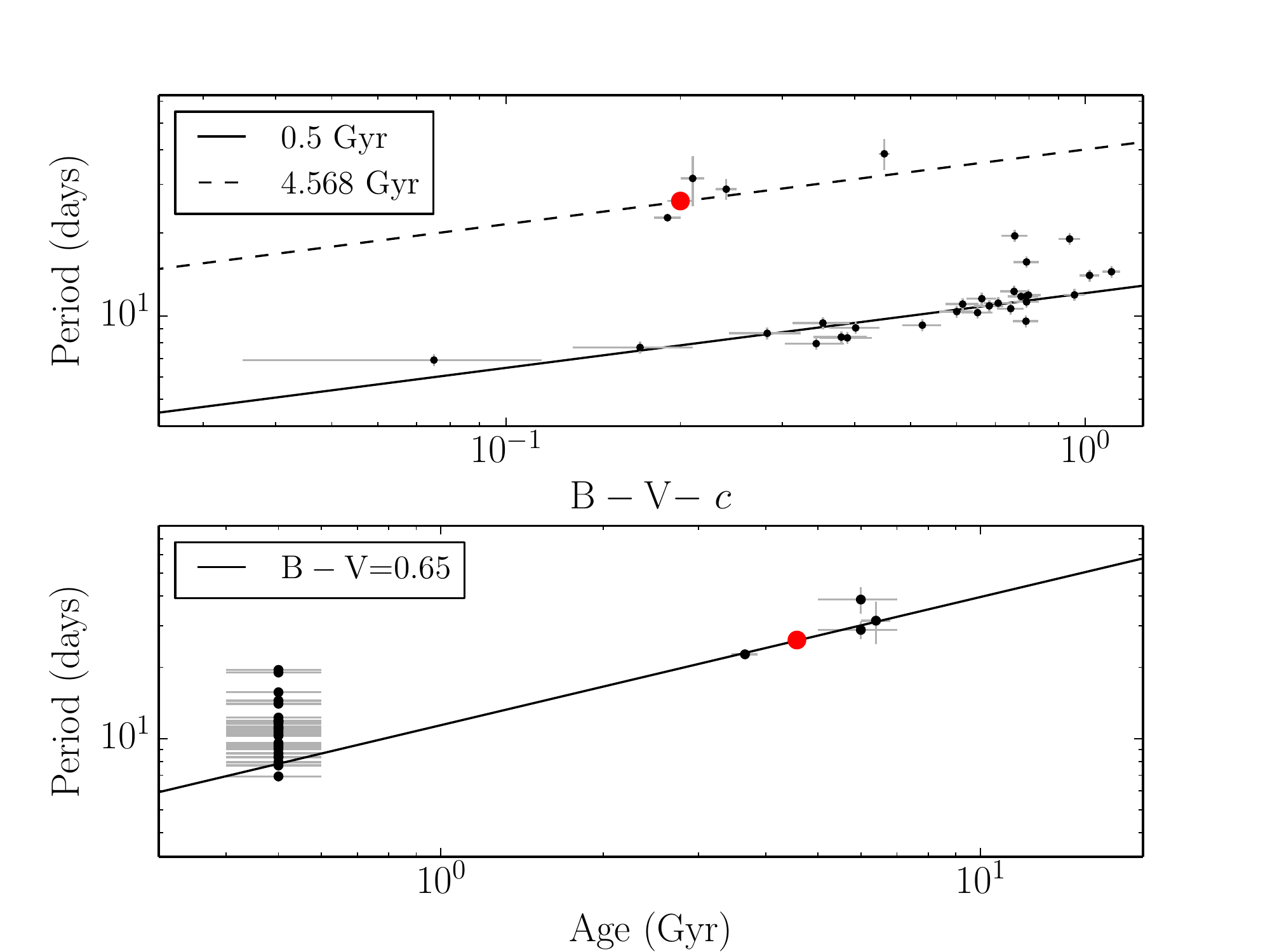}
        }
	\subfigure[Rotation period vs age for the Hyades (top) and Coma
	Berenices (bottom).]{
            \label{fig:HF45}
	    \includegraphics[width=\columnwidth]{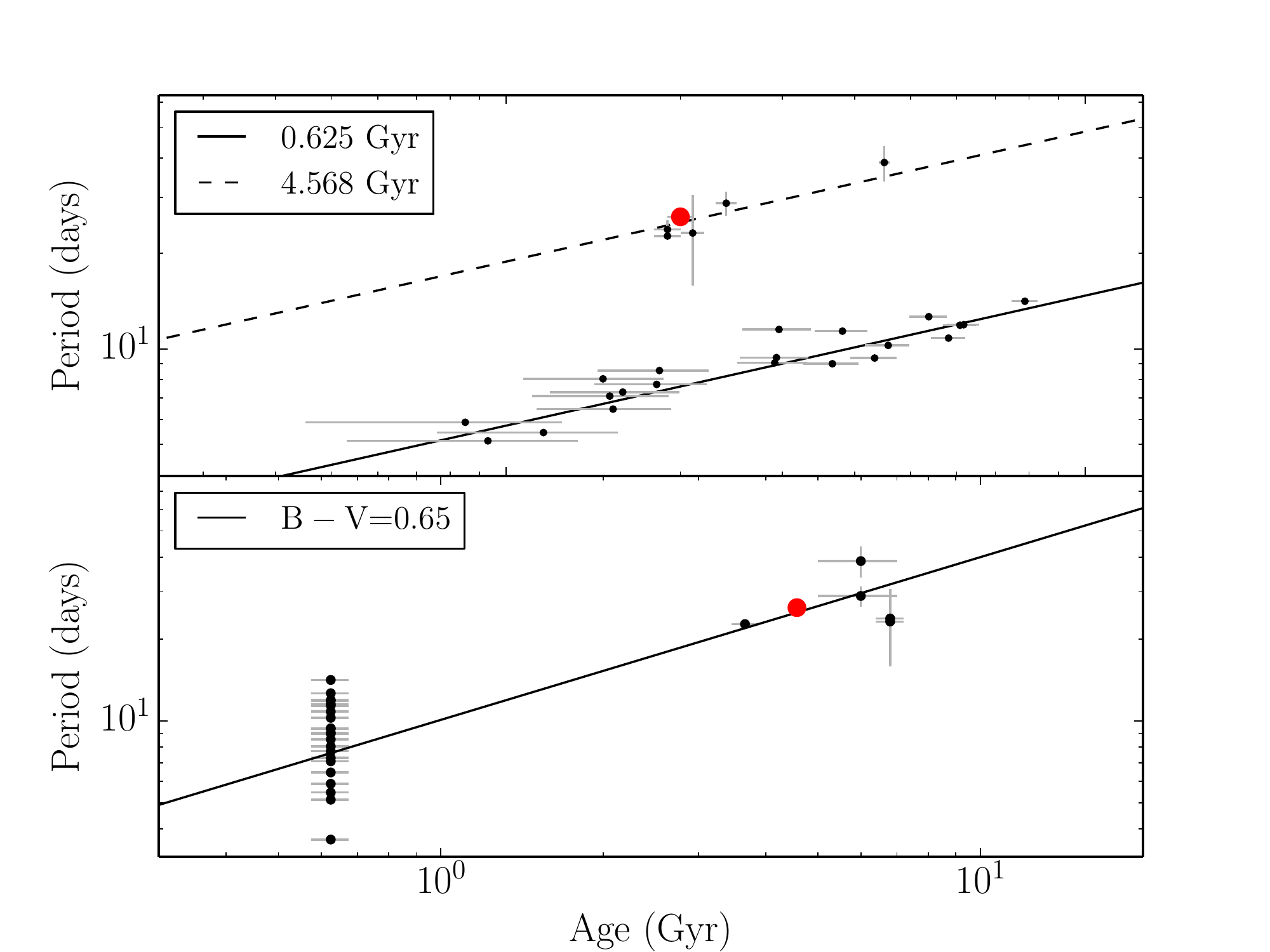}
        }
    \end{center}
    \caption{ Individual fits to the clusters and field stars. The Sun is the
	    red point. The top figure shows rotation period vs `$B-V-c$' for
	    the Hyades (top) and Coma Berenices (bottom) with Solar and cluster
	    age isochrones. The bottom figure shows rotation period vs age
	    for the Hyades (top) and Coma Berenices (bottom) with the period-
	    age relation for a constant $B-V$ value of 0.65 (Solar $B-V$).
\label{fig:subfigures1}}
\end{figure}

A gyrochronology relation was initially fit to the asteroseismic stars, the
field stars and four clusters (Hyades, Coma Berenices, Praesepe and NGC 6811)
all together.
However, this resulted in extremely multi-modal posteriors PDFs for
$a$, $b$ and $n$.
After fitting a separate relation to various subsets of the data, it became
evident that a lower value of $b$ was preferred by NGC 6811,
i.e. the slope of the $\log(\mathrm{period})-\log(B-V-c)$ relation was
shallower for NGC 6811 than for the Hyades, Coma Berenices and Praesepe.
We therefore exluded NGC 6811 from our sample and attempted to fit
a relation to the remaining data.
Multi-modal posterior PDFs were still produced however, until Praesepe was
also removed from the sample.
The reason for this multi-modality is unclear, however we tentatively attribute
it to Praesepe preffering a different value for the colour discontinuity, $c$,
to the Hyades and Coma Berenices.  
We calculated the likelihood for Praesepe, plus the field stars (to provide
the age dependence) with two different values of $c$: $0.45$ and $0.5$,
finding a higher likelihood for $c=0.5$.
Since we do not fully understand the cause of this variation in $c$, and in
order to keep our model simple, we chose to also exclude Praesepe from
our final data set and fit a gyrochronology relation with $c=0.45$ to the
remaining data (asteroseismic stars, field stars, Hyades and Coma Berenices).
We also fitted relations with $c$ values ranging from 0.4 to 0.55 to this final
data set, finding that the results were relatively insensitive to variations
in this parameter (solar ages predicted from each best-fitting model were
consistent within uncertainties).
Individual fits to Hyades and Coma Berenices, plus the field stars are shown
in figures \ref{fig:CF45} and \ref{fig:HF45}.
Median values of $a$, $b$ and $n$ for the two clusters with their 16th and 84th
percentile uncertainties are provided in table
\ref{tab:cluster_results}.
Note that none of these parameters are fully consistent between the two
clusters.
The fact that each cluster seems to prefer a different value of $a$, $b$, $n$
and $c$ paints a concerning picture for this form of a gyrochronology relation
which assumes one set of parameters can be used to describe all stars.
It is likely that the inability of equation \ref{eq:Barnes2007_2} to fully
describe the observed properties of our sample of stars is due to the
simplifying assumptions that go into this relationship.
The fact that there is no dependence on metallicity, for example, may
contribute to the deficiencies of this model.
One might expect metallicity to have an effect on the rotational evolution of
a star since it impacts its internal structure.
We have not attempted to calibrate a metallicity-dependent gyrochronology
relation, because there are no precise metallicity measurements for the
majority of the {\it Kepler} stars, on which this study is chiefly based.
In addition, we anticipate that the majority of stars for which this new
gyrochronology relation will be most useful are {\it Kepler} stars, or stars
targeted by missions like {\it Kepler}, again, the majority of which are unlikely to
have precise metallicities.  

\begin{table}
\caption{Median values of the parameters describing the populations of
	non-gyrochronological stars. \label{tab:nuisance}}

\begin{center}
\begin{tabular}{lcc}
\hline\hline
{Parameter} & {Median value} \\
\hline
$U$ & \U$^{+\Uerrp}_{-\Uerrm}$ days \\
$V$ & \V$^{+\Verrp}_{-\Verrm}$ days \\
$W$ & \W$^{+\Werrp}_{-\Werrm}$ days \\
$X$ & \X$\pm\Xerr$ days \\
$Y$ & \Y$^{+\Yerrp}_{-\Yerrm}$ days \\
$Z$ & \Z$^{+\Zerrp}_{-\Zerrm}$ days \\
$Q$ & \Q$^{+\Qerrp}_{-\Qerrm}$ \\
\hline
\end{tabular}
\end{center}
\end{table}

\begin{figure*}
\begin{center}
\includegraphics[width=6in, clip=false, trim=0 0 0.5in 0]{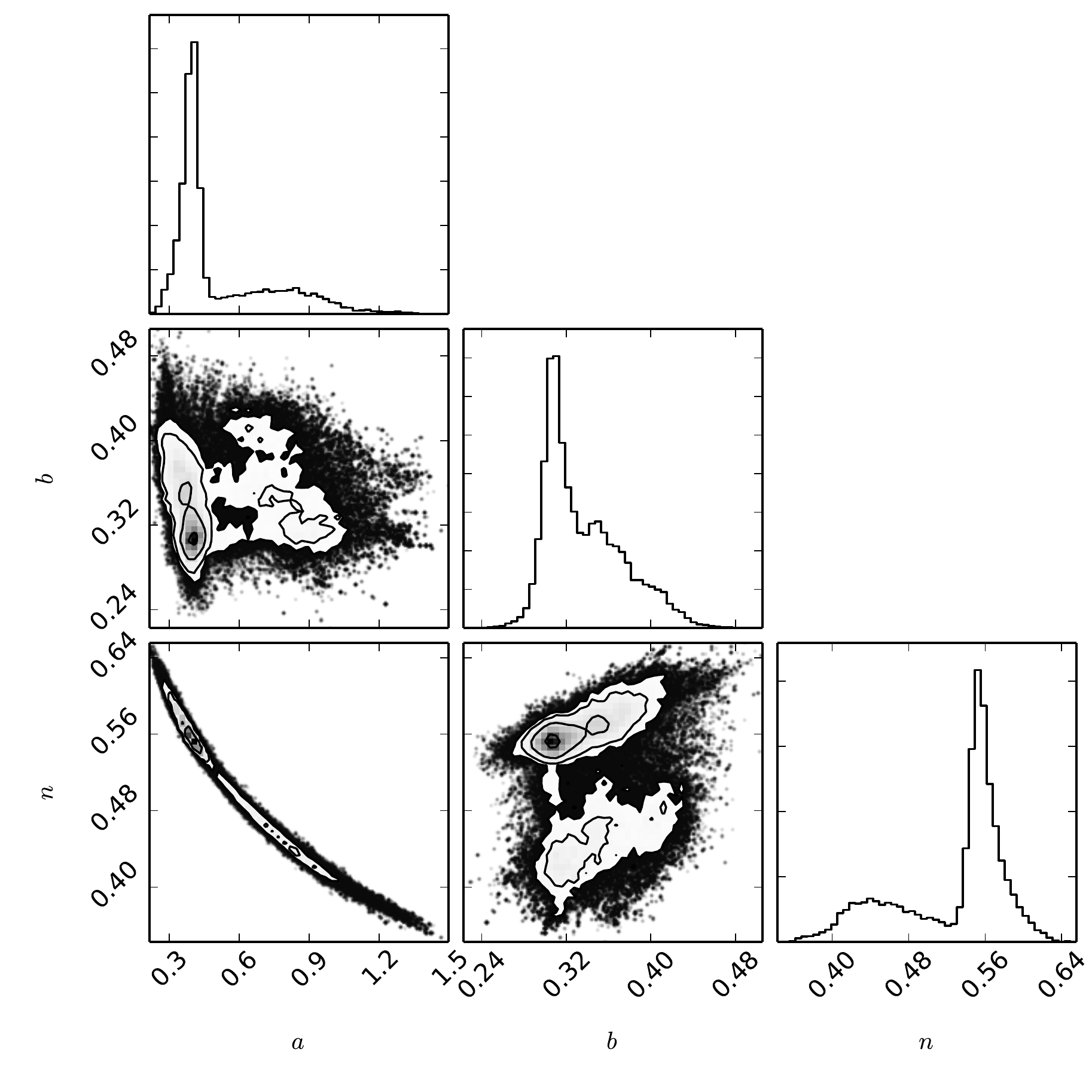}
\caption{Marginalised likelihoods for the three gyrochronology
parameters, $a$, $b$ and $n$. Parameters $a$ and $n$ are correlated and their
posterior PDFs are bimodal. The main peaks in the posterior PDFs of $a$ and
$n$ correspond to a fit to the Sun and field stars. The smaller peak
corresponds to a fit to the {\it Kepler} asteroseismic stars.
This plot was made using triangle.py
\citep{Foreman-Mackey_triangle}.
\label{fig:triangle}}
\end{center}
\end{figure*}

\begin{figure*}
\begin{center}
\includegraphics[width=6in, clip=true, trim=0 0 0.5in 0]{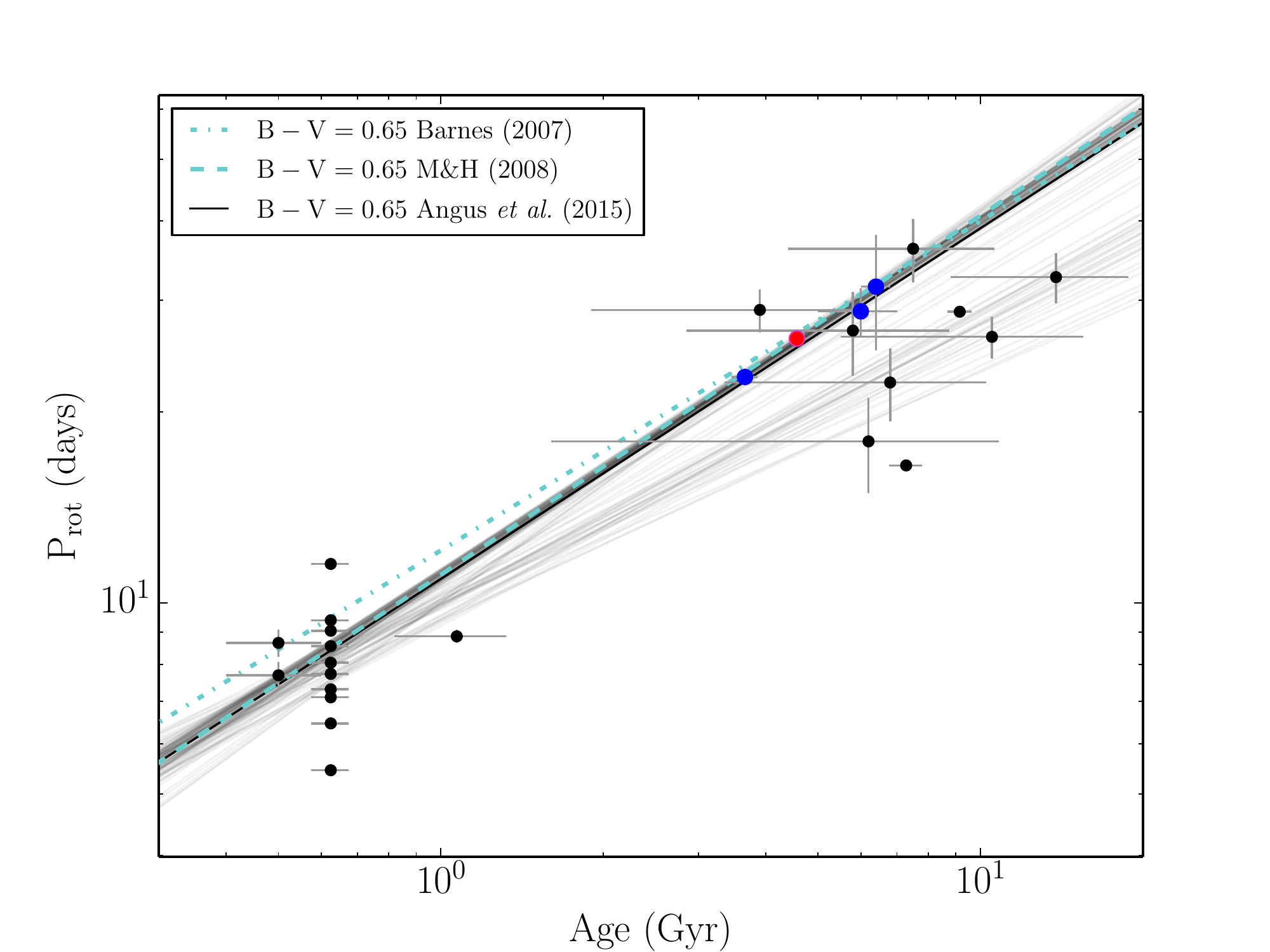}
\caption{Rotation period vs age for cool dwarfs with colour within 0.1 of the
	Sun's: 0.65, with gyrochronology relations of \citet{Barnes2007},
	\citet{Mamajek2008} and this work.
	The Sun is shown in red and the
	field stars, $\alpha$ Cen A, 18 Sco and 16 Cyg B from left to right,
	are shown in blue.
	The black points towards the lower left are cluster stars and those
	towards the upper right are {\it Kepler} asteroseismic stars.
	Each of the faint grey lines represents a
	sample drawn from the posterior probability distributions of $a$, $b$
	and $n$.
	Whilst most of these draws come from the large peak in the posterior
	PDF and fall through the Sun and field stars, some describe the
	period-age relation of the {\it Kepler} asteroseismic stars.
	These lines are drawn from the smaller peak in the posterior PDF.
\label{fig:p_vs_a_solar}}
\end{center}
\end{figure*}

\begin{figure*}
\begin{center}
\includegraphics[width=6in, clip=true, trim=0 0 0.5in 0]{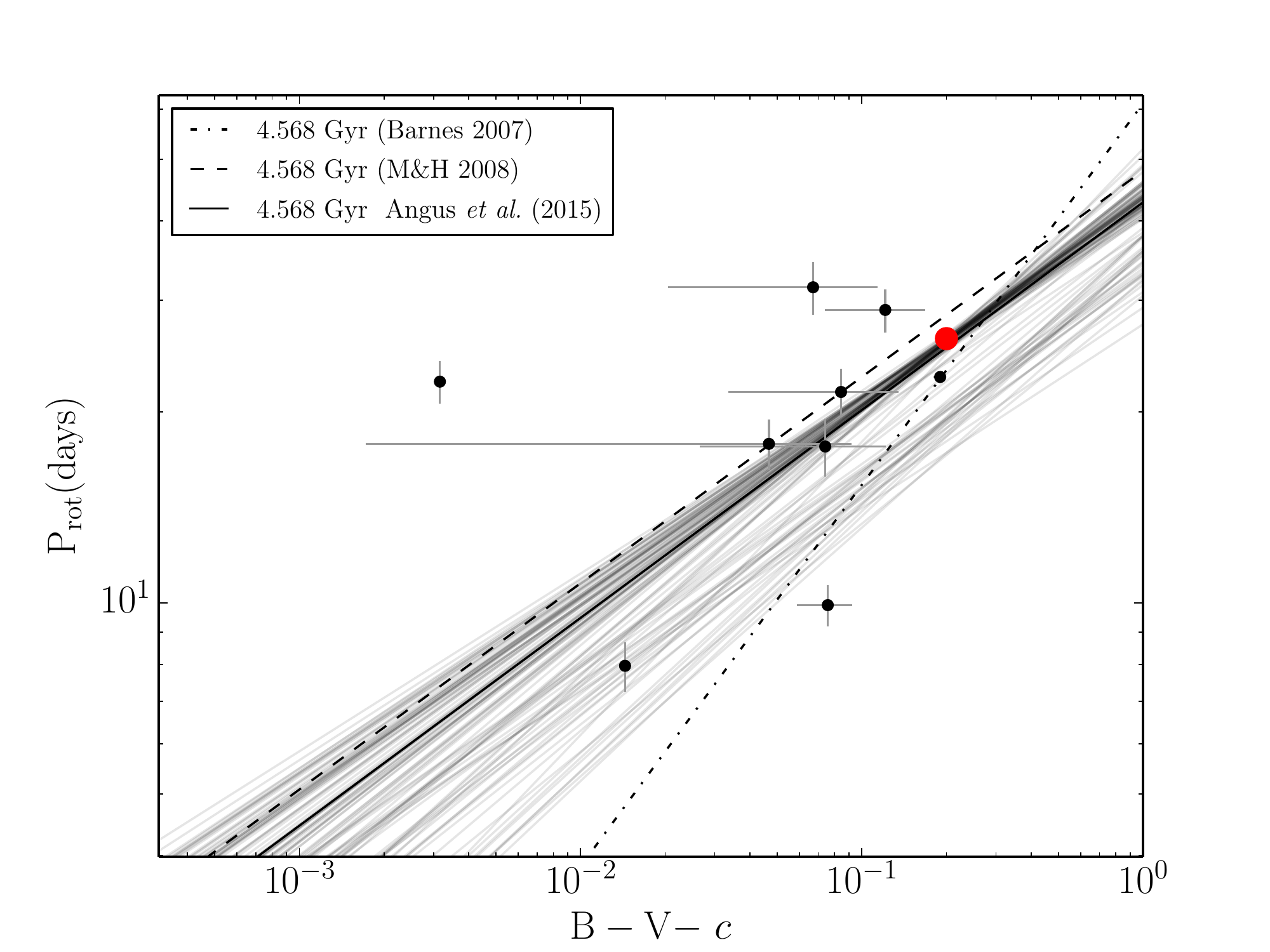}
\caption{Rotation period vs `$B-V-c$' for dwarfs with age within 1$\sigma$ of
	the Sun's age, 4.568 Gyr.
	The Sun is the red point.
	Each of the faint grey lines represents a
	sample drawn from the posterior probability distributions of $a$, $b$ and $n$.
	Many samples fall below the solid black line marking the highest
	probability parameter values due to the bimodal posterior.
\label{fig:p_vs_bv_solar}}
\end{center}
\end{figure*}

\begin{figure*}
\begin{center}
	\subfigure[0.5 Gyr (age of the Coma Ber)]{
            \label{fig:5}
	    \includegraphics[width=\columnwidth]{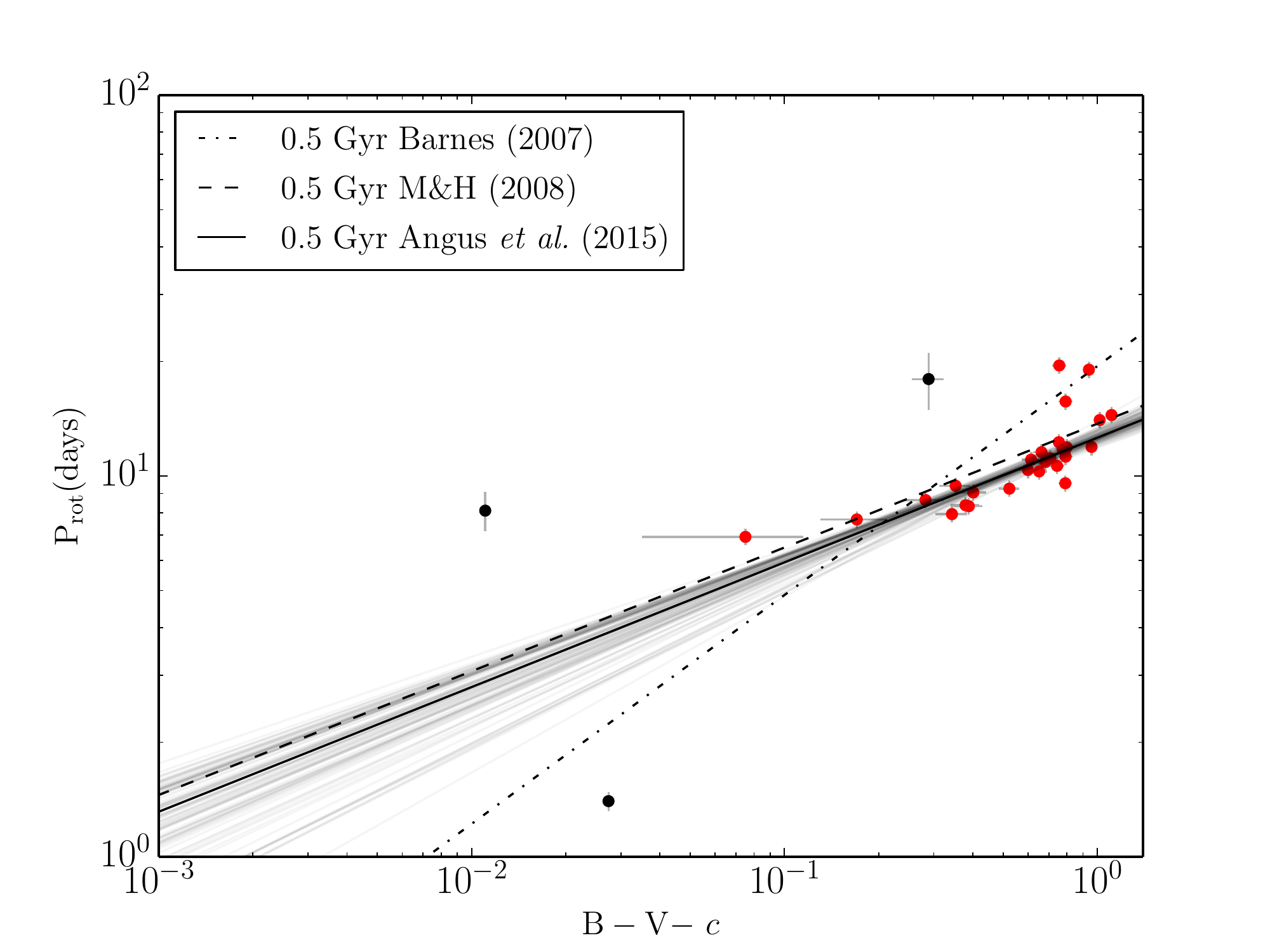}
        }
	\subfigure[0.625 Gyr (age of the Hyades)]{
            \label{fig:625}
	    \includegraphics[width=\columnwidth]{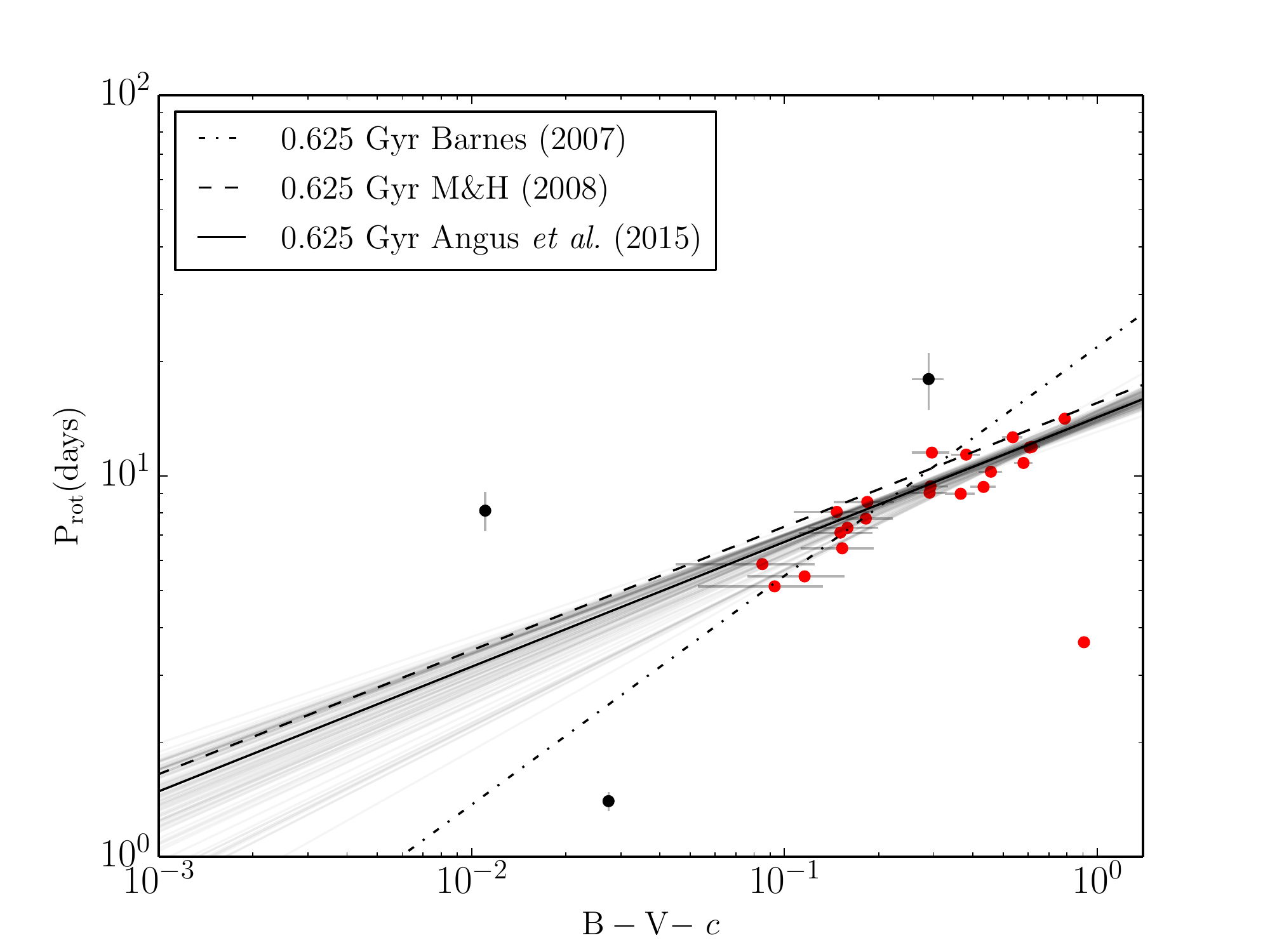}
        }
	\subfigure[2 Gyr]{
            \label{fig:2gyr}
	    \includegraphics[width=\columnwidth]{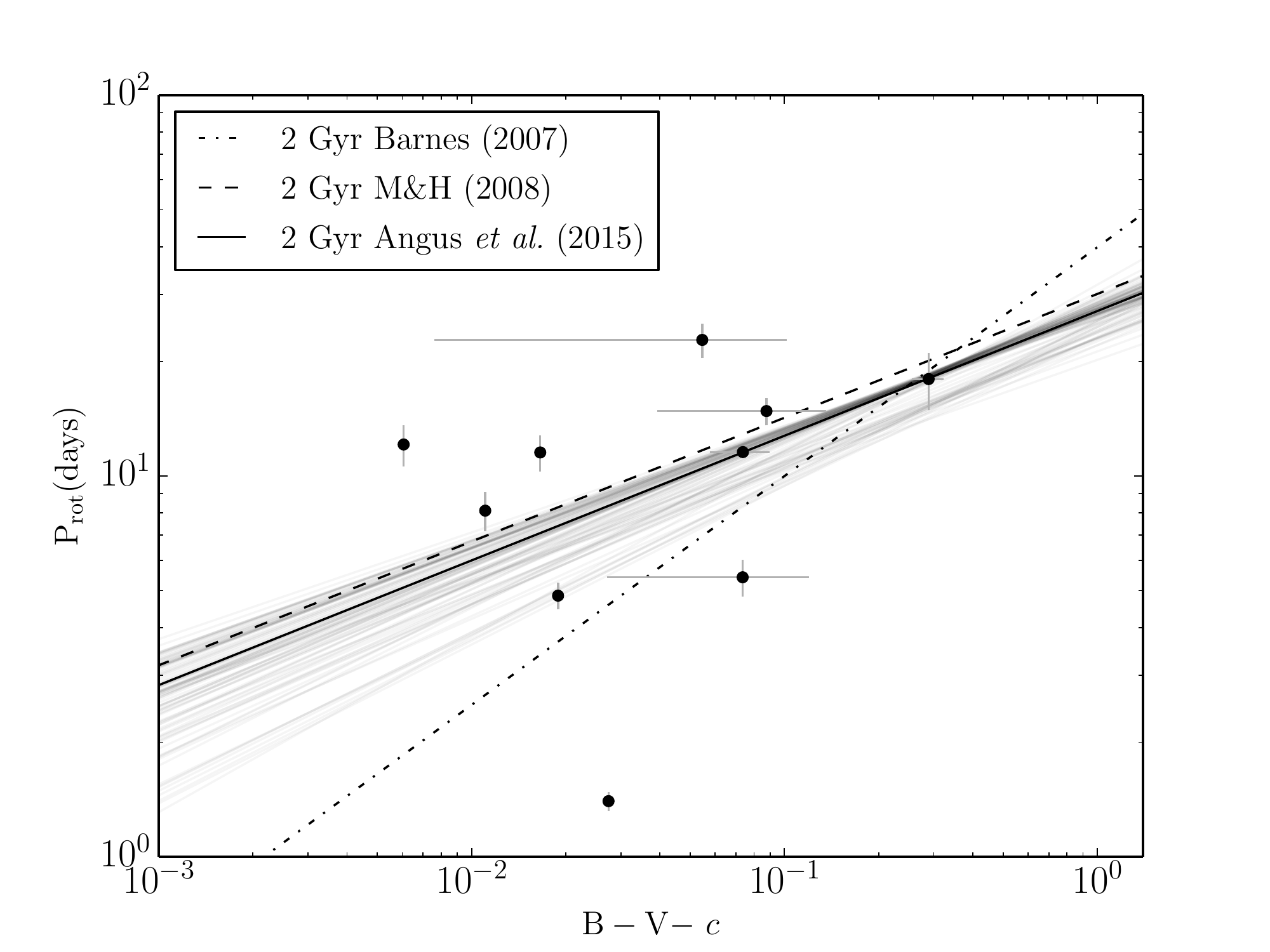}
        }
	\subfigure[5 Gyr]{
            \label{fig:sungyr}
	    \includegraphics[width=\columnwidth]{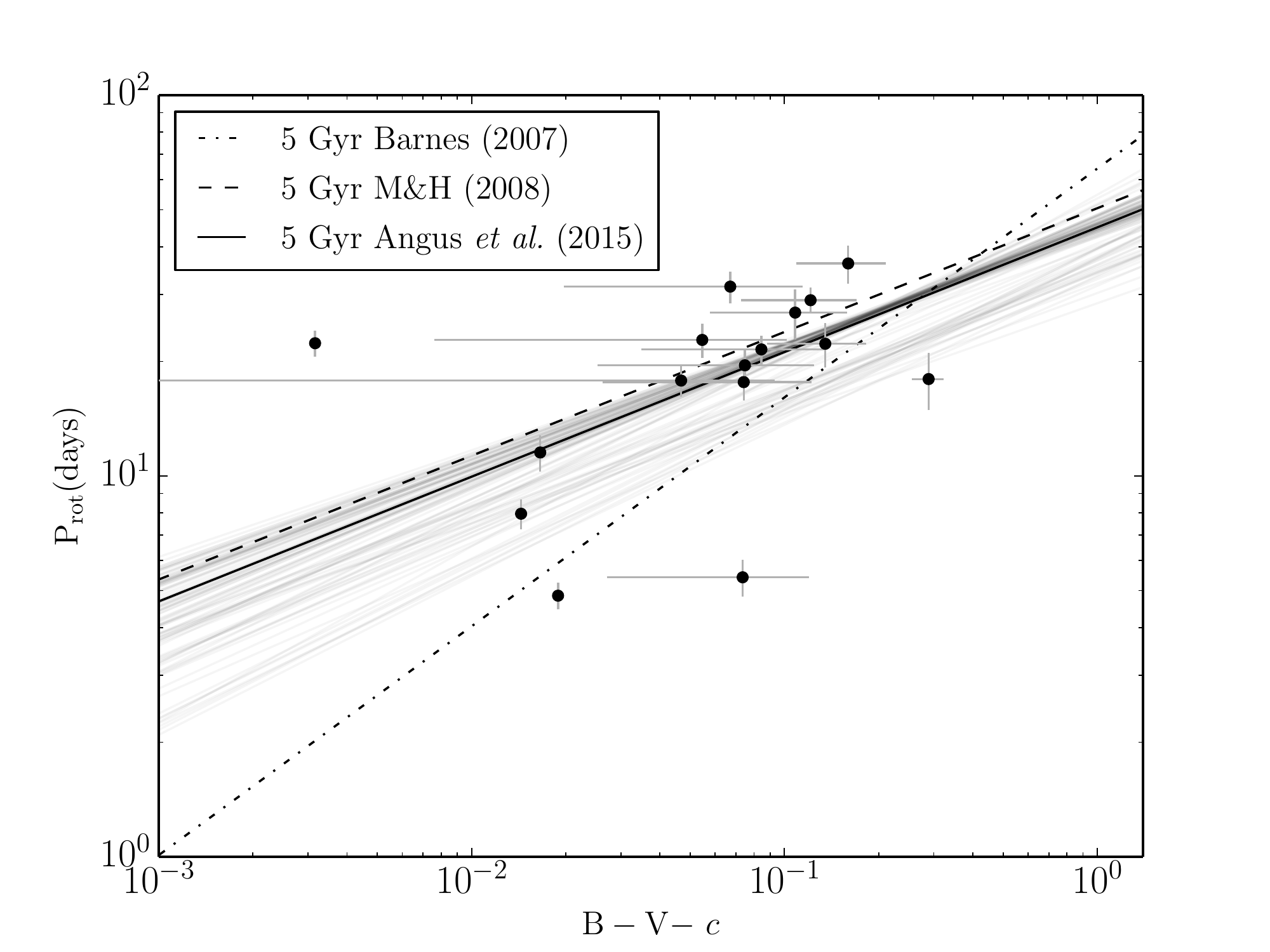}
        }
	\subfigure[8 Gyr]{
            \label{fig:8gyr}
	    \includegraphics[width=\columnwidth]{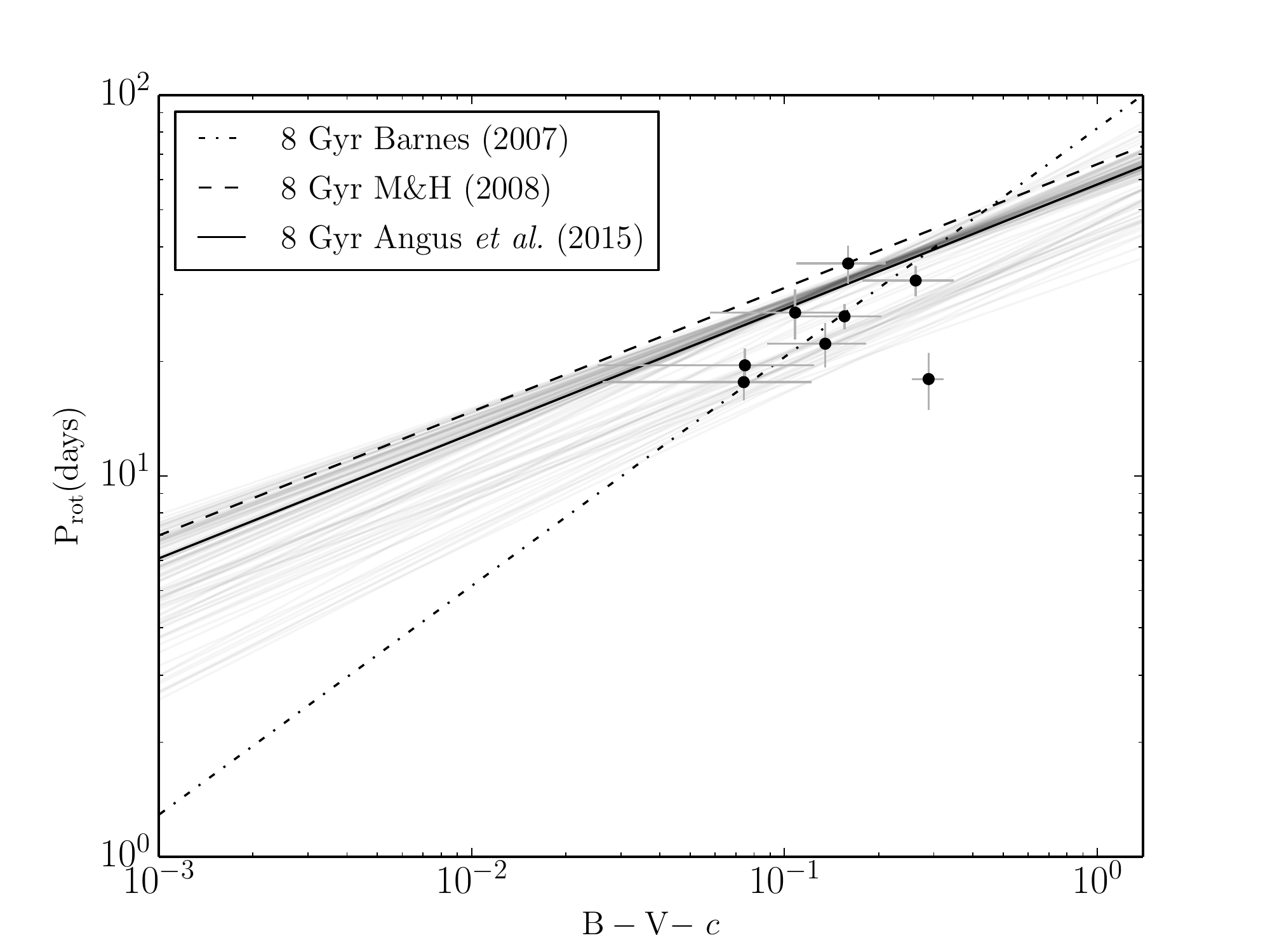}
        }
    \end{center}
    \caption{ \prot vs $B-V-c$ for dwarfs within 1$\sigma$ of the
reference age with the new gyrochronology relation and \citet{Barnes2007}, and
\citet{Mamajek2008} for comparison.
Asteroseismic targets are black and
cluster and field stars are red.
Each of the faint grey lines represents a
sample drawn from the posterior probability PDFs of $a$, $b$ and $n$.
Many samples fall below the best fit model due to the bimodal posterior PDFs.
\label{fig:subfigures2}}
\end{figure*}

The resulting highest probability values of $a$, $b$ and $n$ from our fit to
the final data set, with 16th and 84th percentile uncertainties are
presented in table \ref{tab:constants}.
The additional parameters of our model, describing the distributions of hot
star and subgiant rotation periods, are presented in table \ref{tab:nuisance}.
The posterior PDFs of these parameters were all unimodal.
Note the value of $Q$, the parameter describing the fraction of misclassified
subgiants is $\Q^{+\Qerrp}_{-\Qerrm}$, i.e., based on our simple `\logg$=4.2$'
definition of MS turn-off, which left only 21 stars classified as cool dwarfs,
one or two of these are likely to be misclassified subgiants.
Marginalised posterior PDFs for the three gyrochronology parameters are shown
in figure \ref{fig:triangle} and the resulting relation between period and age
for stars of Solar-like colour is shown in figure \ref{fig:p_vs_a_solar}.
The relation between period and colour for Solar-age stars is shown in figure
\ref{fig:p_vs_bv_solar} and for a range of ages in figures
\ref{fig:5}-\ref{fig:8gyr}.
Note that we plot rotation period vs `$B-V-c$', producing a straight line, in
order to give a more intuitive understanding of the quality of the fit to the
data.
In figures \ref{fig:p_vs_a_solar} to \ref{fig:8gyr} we plotted 100 draws from
the posterior PDFs of the gyrochronology parameters as faint grey lines,
in order to demonstrate the widths and bimodal natures of these distributions.

Figure \ref{fig:triangle} shows that parameters $a$ and $n$ are
correlated and their posterior PDFs are bimodal.
The position of the second peak falls around $a = 0.8$, $b = 0.34$ and
$n = 0.44$.
The cause of this bimodality is clear when looking at the faint grey lines
representing draws from the parameter posterior PDFs in
figure \ref{fig:p_vs_a_solar}.
The majority of these draws fall close to the best-fitting model, which passes
neatly through the Sun and field stars, however a significant fraction fall
below the line of best-fit, passing through the {\it Kepler} asteroseismic stars
which also mostly fall below the line.
This result is reflected in \citet{Garcia2014} who model the AMP
{\it Kepler} asteroseismic stars from \citet{Metcalfe2014}, without anchoring their
relation to the Sun.
They find that the model that best describes the {\it Kepler} asteroseismic stars
underpredicts the rotation period of the Sun.
The grey lines that fall beneath our best-fitting model in figure
\ref{fig:p_vs_a_solar} are drawn from the smaller peak in the posterior PDFs
of $a$ and $n$ and seem to describe the relation between rotation period and
age for the {\it Kepler} asteroseismic stars.
In other words, the bimodal posterior PDFs of $a$ and $n$ are produced by the
disagreement between the {\it Kepler} asteroseismic stars and the Sun and field
stars.
One set of gyrochronology parameters is not
capable of describing the Sun, plus field stars, and the {\it Kepler}
asteroseismic stars simultaneously.
There is more than one possible explanation for this result.
Firstly, the asteroseismic ages could be systematically biased high.
Secondly, the rotation periods could be systematically underestimated.
This could occur if, for example, these stars were rotating differentially
and the dominant spotted regions on their surfaces were not equatorial.
Thirdly, this could be a result of an observational bias produced by incomplete
detection, sometimes known as Malmquist bias \citep{Malmquist1920}.
If there were a large spread in rotation periods for a given stellar mass and
age and, due to the detection bias brought about because shorter periods are
easier to detect than longer periods, this broad range of rotation periods
might be truncated at some upper cut-off.
It would therefore appear as though Solar-colour stars were rotating too
slowly for their age simply because only rapidly rotating {\it Kepler} stars
appeared in our sample.
{\it Kepler} systematics hinder our ability to measure longer rotation periods,
but in addition, more slowly rotating stars tend to be less active, with
fewer surface features and are therefore more likely to be missing from our
sample.
Finally, it is possible that the {\it Kepler} asteroseismic stars follow a
\emph{different} spin-down relation to the Sun and field stars, perhaps due to
having different metallicities. 
Unfortunately it is not currently possible to identify the cause of the
observed mismatch between the {\it Kepler} stars and the field stars and we leave
this question for a future investigation.

The fact that our fit is so heavily dominated by the Sun with its small
uncertainties is, perhaps, a cause for concern.
We know the rotation period and age of the Sun very precisely, however, if the
Sun is not \emph{exactly} representative of a typical star, the resulting
best-fit gyrochronology relation will also not represent typical stars.
The field stars in our sample \emph{are} well represented by the best-fit
gyrochronology relation, which provides reassurance that the Sun is a typical
star amongst this set.
We did not attempt to tackle the problem of how to appropriately treat the Sun
as a single, highly-precise data point in a sample that also contains imprecise
data.
Instead, we leave this problem for future consideration.
Despite the fact that a lot of weight is attributed to the Sun because of its
small uncertainties, this new gyrochronology relation is still the most
representative, empirically calibrated relation between colour, rotation period
and age for MS F, G and K stars to date. 

Our final, newly calibrated gyrochronology relation can be written in full as
\begin{equation}
	P = A^{\gyron^{+\nerrp}_{-\nerrm}} \times \gyroa^{+\aerrp}_{-\aerrm}
	(B-V-0.45)^{\gyrob^{+\berrp}_{-\berrm}},
\label{eq:Barnes2007_3}
\end{equation}
with rotation period, $P$ in days and age, $A$ in Myr.
An age can be calculated for a star with a rotation period and colour by
inverting this relationship.
Covariances between the gyrochronology parameters should
be taken into account {\it whenever} the above relation is used to calculate
uncertainties on an age or rotation period prediction.
In order to do this properly, posterior PDF samples should be incorporated
into Monte-Carlo uncertainty calculations\footnote{Posterior samples for
this project are available at https://github.com/RuthAngus/Gyro.}.

In order to test the predictive power of the new gyrochronology relation, we
inverted equation \ref{eq:Barnes2007_3} to compare previously measured ages
with new age predictions for the 6 field stars (see table
\ref{tab:comparison}).
For comparison, gyrochronological ages for the field stars were also computed
using the relations of \citet{Barnes2007} and \citet{Mamajek2008}.
Uncertainties on ages predicted with the new relation were calculated using
posterior PDF samples of the three parameters, $a$, $b$ and $n$.

\begin{table*}
\caption{Field star ages taken from the literature, compared with
	predictions from this work (1), \citet{Mamajek2008} (2)
	and \citet{Barnes2007} (3). \label{tab:comparison}}

\begin{tabular}{lcccc}
\hline\hline
{Star} & {Literature age (Gyr)} & {Age 1 (Gyr)} & {Age 2 (Gyr)} & {Age 3 (Gyr)} \\
\hline

18 Sco      & $3.7 \pm 0.2$     & $3.7^{+2.5}_{-0.3}$ & $3.5^{+0.6}_{-0.5}$
	    & $3.7^{+0.8}_{-0.6}$ \\

The Sun     & $4.568 \pm 0.001$ & $4.6^{+3.5}_{-0.3}$ & $4.7^{+0.7}_{-0.6}$
	    & $4.8^{+1}_{-0.8}$ \\

Alpha Cen A & $6.0 \pm 1$       & $5.0^{+3.3}_{-1.0}$   & $4.5^{+1}_{-0.9}$
	    & $5\pm1$ \\

Alpha Cen B & $6.0 \pm 1$       & $6.0^{+3.8}_{-1.7}$       & $4^{+1}_{-0.9}$
	    & $5^{+2}_{-1}$ \\

16 Cyg A    & $6.8 \pm 0.4$     & $4.0^{+2.6}_{-0.7}$       & $3\pm2$
	    & $4\pm2$ \\

16 Cyg B    & $6.8 \pm 0.4$     & $3.7^{+3.4}_{-2.1}$       & $3\pm2$
	    & $4\pm2$ \\
\hline
\end{tabular}
\end{table*}

The ages predicted by the three different relations are consistent within
uncertainties, with the exception of Alpha Cen B.
All three relations underpredict the age of the 16 Cyg system.
The rotation periods of both 16 Cyg A and B used in this paper are
asteroseismic measurements of the internal (not surface) rotation periods
of the stars.
If the stars' cores are rotating much more rapidly than their surfaces, this
could account for this age discrepancy.

The goal of gyrochronology in general is to provide a means of predicting the
age of a star given observations of its colour (or mass, or temperature), and
rotation period.
The discrepancies in period-colour relations between clusters, {\it Kepler} stars
and nearby field stars in the above analysis does not bode well for the
current, simple gyrochronology model.
Until now it has been hoped that one single relation between period, mass and
age could be used to describe all F, G and K MS stars.
Even though we are unable to identify the cause of the discrepancy between the
{\it Kepler} stars and the field stars, a different gyrochronology relation still
seems to be required for each cluster.

The `narrowness' of the gyrochronology relation has hitherto been an unknown;
do the three properties, age, mass and rotation period, truly lie on an
infinitely narrow plane?
Unfortunately we cannot fully answer this question here as the asteroseismic
ages are noisy and observational and intrinsic scatter are ambiguously
interwoven.
However, if this sample of stars is subject to Malmquist bias then, by
definition, there must be a broad range of rotation periods for stars of a
given mass and age.
A future study might include an extra parameter that describes the `width' of
the gyrochronological plane and attempt to detect an element of scatter above
the noise level.
Does age depend solely on rotation period and mass or do other variables
influence stellar spin down, perhaps only becoming important after many Myrs?
The gyrochronology model calibrated here neglects the effects of metallicity.
This property is bound to have an effect on the angular momentum evolution
of a star since it impacts internal stellar structure.
A future study, investigating the impact of metallicity on rotational evolution
will be essential for improving our ability to date stars using gyrochronology.

The picture of gyrochronology will become clearer as the sample of
asteroseismic stars with individual mode analysis grows and their age
uncertainties shrink.
The best targets for asteroseismic studies are relatively inactive since they
allow the easier detection of Solar-like oscillations.
Inactive stars are also the best targets for gyrochronology as they are
usually old and slowly rotating.
However these targets are not well suited for
rotation period measurements which are most easily and precisely determined
for active, rapidly rotating stars.
K2, the repurposed {\it Kepler} mission, will provide new targets for rotation
studies; in particular, some relatively old clusters have been and will
continue to be monitored by the spacecraft.
The observing seasons of K2 are relatively short ($\sim$ 90 days) and fields
will only be observed once, so the maximum rotation periods measurable from
K2 light curves will be considerably shorter than with {\it Kepler}.
However these clusters may still be extremely useful for gyrochronology.

\section{Summary}
\label{sec:conclusions}

We have calibrated the relation between rotation period, $B-V$ colour and age
for MS stars with \teff$<$ 6250 K, using \nastero$~${\it Kepler} asteroseismic
targets, supplemented with 6 field stars and \nHC$~$cluster stars.
Unlike previous gyrochronology calibrations, our sample covers a large range
of ages, observational uncertainties on all parameters were accounted for and
the posterior PDFs of model parameters were explored using MCMC.
Incorporating observational uncertainties into the model fitting process was
an essential component of our analysis since these uncertainties, particularly
on the asteroseismic ages, were large.
Three populations: hot dwarfs, cool dwarfs and subgiants, were modelled
simultaneously in order to account for potential misclassifications that
might have arisen from large observational uncertainties.
Posterior probability distributions of the gyrochronology parameters were
explored using MCMC and the impact of leaving out subsets of the sample was
assessed, leading us to find that a single relation between rotation period,
colour and age does not adequately describe the cluster data.
For this reason, only the Hyades and Coma Berenices clusters were used to
calibrate the model.
Fitting equation \ref{eq:Barnes2007_2} to the Hyades and Coma Berenices
clusters, the {\it Kepler} asteroseismic stars, local field stars and the Sun
resulted in bimodal posterior PDFs.
The {\it Kepler} asteroseismic stars were not well described by the same
gyrochronology relation as the Sun, field stars and clusters.
There are several potential explanations for the cause of this phenomenon.
Some of the most likely are: the asteroseismic ages or the photometric
rotation periods could be systematically biased; the rotation periods might be
subject to Malmquist bias due to incomplete detection; and the different
populations could have different physical properties (e.g. metallicity), not
accounted for in the gyrochronology model.
If Malmquist bias is responsible and there truely {\it is} a broad range of
rotation periods for stars of a given age and colour, this still has
negative implications for Gyrochronology which relies on the assumption that
these three properties lie on a neat plane.


The careful and well motivated modelling techniques used in this work
allowed us to identify the potential shortcomings of the current
gyrochronology model.
Only by conducting MCMC parameter exploration and studying the resulting
posterior PDFs were we able to see that one global gyrochronology relation
cannot describe all subsets of our sample.
The results are somewhat unsettling---no single model describes the sample as
a whole, even when dropping some of the clusters and allowing for rather
complex population subsets.
This tells us that the current model is probably not good enough, but it - or
something very similar - is the model everyone has been using in
gyrochronology studies so far.
Previous studies did not model multiple cluster and field populations
simultaneously, or do the modelling in such a detailed way, so the problems
didn’t come through so clearly.
In the future better physical models may be developed, better ages may be
calculated and more sensitive period search methods, not so susceptible to
Malmquist bias may become available.
When they do, the prescription we provide will enable the the
period-colour-age relation to be modelled more sensitively.


The code used in this project can be found at
https://github.com/RuthAngus/Gyro.
R.A. acknowledges funding from the Leverhulme Trust.
A.M. acknowledges funding from the European Research Council under the EU’s
Seventh Framework Programme (FP7/(2007-2013)/ERC Grant Agreement No. 291352).
We wish to thank the anonymous referee for their valuable and insightful
feedback and Bill Chaplin for his thoughtful and detailed comments
which enormously improved this paper.
We would also like to thank Tsevi Maseh, Eric Mamajek, Daniel Mortlock and
David Hogg for useful insight and discussion that hugely contributed to this
paper.

\bibliographystyle{mn2e}
\bibliography{Gyro_paper}

\end{document}